\documentclass{article}

\usepackage{arxiv}

\usepackage[utf8]{inputenc} % allow utf-8 input
\usepackage[T1]{fontenc}    % use 8-bit T1 fonts
\usepackage{hyperref}       % hyperlinks
\usepackage{url}            % simple URL typesetting
\usepackage{booktabs}       % professional-quality tables
\usepackage{amsfonts}       % blackboard math symbols
\usepackage{nicefrac}       % compact symbols for 1/2, etc.
\usepackage{microtype}      % microtypography
\usepackage{lipsum}		% Can be removed after putting your text content
\usepackage{graphicx}
\usepackage{natbib}
\usepackage{doi}

\usepackage{wrapfig}
\usepackage{cleveref}
\usepackage{bm}
\usepackage{soul}
\usepackage{tikz}
\usepackage{threeparttable}

\usepackage{titlesec}
\titlespacing*{\subsubsection}{0pt}{0\baselineskip}{0\baselineskip}

\title{KGML-xDTD: A Knowledge Graph-based Machine Learning Framework for Drug Treatment Prediction and Mechanism Description}

\date{} 					% Or removing it

\author{
    Chunyu Ma \\
    Huck Institutes of the Life Sciences\\ 
    Pennsylvania State University\\
	State College, PA 16802, USA\\
    \texttt{cqm5886@psu.edu} \\
    \And
	Zhihan Zhou \\
	Department of Computer Science\\ 
    Northwestern University\\
    Evanston, IL 60208, USA\\
    \texttt{zhihanzhou@u.northwestern.edu} \\
    \AND
	Han Liu \\
	Department of Computer Science\\ 
    Northwestern University\\
    Evanston, IL 60208, USA\\
    \texttt{hanliu@northwestern.edu} \\
	\And
    David Koslicki \\
    Department of Computer Science\\
    Department of Biology\\
    Huck Institutes of the Life Sciences\\
    Pennsylvania State University\\
	State College, PA 16802, USA\\
    \texttt{dmk333@psu.edu}\\
}

\begin{document}
\maketitle

\begin{abstract}
\textbf{Background}: Computational drug repurposing is a cost- and time-efficient approach that aims to identify new therapeutic targets or diseases (indications) of existing drugs/compounds. It is especially critical for emerging and/or orphan diseases due to its cheaper investment and shorter research cycle compared with traditional wet-lab drug discovery approaches. However, the underlying mechanisms of action (MOAs) between repurposed drugs and their target diseases remain largely unknown, which is still a main obstacle for computational drug repurposing methods to be widely adopted in clinical settings. \\
\textbf{Results}: In this work, we propose \texttt{KGML-xDTD}: a \textbf{K}nowledge \textbf{G}raph-based \textbf{M}achine \textbf{L}earning framework for e\textbf{x}plainably predicting \textbf{D}rugs \textbf{T}reating \textbf{D}iseases. It is a two-module framework that not only predicts the treatment probabilities between drugs/compounds and diseases but also biologically explains them via knowledge graph (KG) path-based, testable mechanisms of action (MOAs). We leverage knowledge-and-publication based information to extract biologically meaningful "demonstration paths" as the intermediate guidance in the Graph-based Reinforcement Learning (GRL) path-finding process. Comprehensive experiments and case study analyses show that the proposed framework can achieve state-of-the-art performance in both predictions of drug repurposing and recapitulation of human-curated drug MOA paths. \\
\textbf{Conclusions}: 
\texttt{KGML-xDTD} is the first model framework that can offer KG-path explanations for drug repurposing predictions by leveraging the combination of prediction outcomes and existing biological knowledge and publications. We believe it can effectively reduce "black-box" concerns and increase prediction confidence for drug repurposing based on predicted path-based explanations, and further accelerate the process of drug discovery for emerging diseases.
\end{abstract}

% keywords can be removed

\section{Introduction}\label{sec:introduction}

Traditional drug development is a time-consuming process (from initial chemical identification to clinical trials and finally to FDA approval) that takes around 10-15 years and also comes along with billions-of-dollars investments and high failure rates \citep{Berdigaliyev_Nurken_2020}. Considering the rapid pace of novel disease evolution, it is urgent to find a more efficient and economical drug discovery method. Fortunately, it has been observed that a single drug can often be effective in treating multiple diseases. For example, thalidomide was originally used as an anti-anxiety medication \citep{Miller_M_T_1991}, and was later found to have the anti-cancer potential for the treatment of cancers \citep{Verheul_H_M_W_1999,Singhal_Seema_1999}. Hence, drug repurposing, also known as the identification of new uses for existing drugs/compounds, might bring us hope to address this urgent need with the advantage of a shorter research cycle, lower development cost, and more preexisting safety tests.

Existing drug repurposing approaches can roughly be categorized into experimental-based approaches (e.g., binding affinity assays \citep{Kairys_Visvaldas_2019}, phenotypic screening \citep{Aulner_Nathalie_2019}), clinical-based approaches (e.g., off-label drug use analysis \citep{Rusz_Carmen_2021}), and computational-based approaches (e.g., chemical-structure-based \citep{Swamidass_S_Joshua_2011}, and GWAS-based approaches \citep{Sanseau_Philippe_2012}). Compared with the former two approaches, the computational approaches are more cost- and time-efficient, particularly when the goal is to prioritize a large number of target drugs/compounds for follow-up experimental investigation. Among all computational drug repurposing methods, the integration of multiple biomedical data sources into a so-called \textbf{biomedical knowledge graph} (BKG) for drug discovery has become popular in recent year \citep{Stephen_Bonner_2022} due to the increasing availability of curated biomedical databases such as DrugBank \citep{Wishart_David_S_2017DrugBank}, ChEMBL \citep{Gaulton_Anna_2012}, HMDB \citep{Wishart_David_S_2017HMDB} and the advancement of semantic web techniques \citep{Samantha_Kanza_2021}. There are three types of existing BKGs: database-based BKGs, literature-based BKGs, and mixed BKGs. The database-based BKGs (e.g., \emph{Hetionet} \citep{Himmelstein_Daniel_Scott_2017}, \emph{BioKG} \citep{Walsh_Brian_2020}, \emph{CBKH} \citep{Su_Chang_2021}) are constructed by integrating biomedical data and their relations stored in existing biological databases. The literature-based BKGs (e.g., \emph{GNBR} \citep{Percha_Bethany_2018}) are built by leveraging Natural Language Processing (NLP) techniques to extract semantic information from a large amount of available biomedical literature and electronic health record (EHR) data, which are mostly disease-specific \citep{Zhang_Rui_2021, Wang_Qingyun_2021, Li_Nan_2020}. The mixed BKGs (e.g., \emph{CKG} \citep{Santos_Alberto_2022}, \emph{RTX-KG2} \citep{Wood_E_C_2022}) are generated by combining the knowledge sources from the above two methods.

Based on these BKGs, several machine learning methods have been proposed or implemented for drug repurposing prediction by treating it as a link prediction task in the BKGs. For example, \citet{Himmelstein_Daniel_Scott_2017} used the so-called degree-weighted path count (DWPC) to assess the prevalence of 1,206 metapaths and then classified drug-disease treatment relations by fitting these DWPC features to a logistic regression model. \citet{Vassilis_N_Ioannidis_2020} proposed a novel graph neural network model I-RGCN to learn the node and relation embeddings for the Covid-19 drug repurposing task. \citet{Zhang_Rui_2021} recently predicted the possible drugs for Covid-19 with five existing popular knowledge graph completion methods (e.g. TransE \citep{Bordes_Antoine_2013}, RotatE \citep{Sun_Zhiqing_2019}, DistMult \citep{Yang_Bishan_2014}, ComplEx \citep{Trouillon_Theo_2016}, and STELP \citep{Wang_Bo_2021}). Although some of these models have shown good performance in drug repurposing prediction on the small-scale BKGs, none have been scaled to massive BKGs with more than millions of nodes and edges and make a comprehensive comparison. More importantly, most of them lack the biological explanatory ability for their predictions, which limits their applicability in clinical research. 

Currently, there are few computational models designed for drug repurposing \textit{explanations}. A common and intuitive explanation based on a biomedical knowledge graph for drug repurposing leverages the semantic BKG-based paths between given drug-disease pairs. \citet{Sosa_Daniel_N_2020} applied a graph embedding model UKGE \citep{Chen_Xuelu_2019}, which utilizes the weighted (the frequency of relation appeared in literature) relation edges in a literature-based KG \emph{GNBR}, to identify new indications of drugs for rare diseases and then explain the results via the highest-ranking paths based on confidence scores. However, this method is only applicable in the literature-based BKGs with the weighted edge information. Most BKGs using database-based knowledge don't contain such information. \citet{Sang_Shengtian_2019} proposed GrEDeL that combines the TransE embedding method with a Long Short-Term Memory (LSTM) Recurrent Neural Network (RNN) model to predict drug-disease relation. By using the embeddings of BKG-paths as model input for predictions, they can provide path-based explanations. However, they claimed that the effectiveness of the approach relies heavily on the NLP tool SemRep, which is reported to have high false positives in named entity recognition \citep{Kilicoglu_Halil_2020}. Also, they didn't fully evaluate how biologically reasonable their predicted path-based mechanisms of action (MOAs) are. 

Besides the existing methods above, we view reinforcement learning (RL) as a promising solution for drug repurposing explanation. RL models solve the decision-making problem, in which an agent learns how to take appropriate actions to maximize cumulative rewards through interactions with the environment. RL has achieved widespread success in various domains, including games, recommendation systems, healthcare, transportation, etc \citep{Li_Yuxi_2019}. Graph Reinforcement Learning (GRL), first proposed around in 2017, aims to solve graph mining tasks such as link prediction \citep{Chen_Ling_2022}, adversarial attacks \citep{Sun_Yiwei_2020}, and relational reasoning \citep{Zhou_Xingchen_2021}. Unlike its applications in other domains, one of the biggest challenges in GRL is finding an appropriate reward to guide the path searching in specific domains. To address the issue of finding biologically reasonable BKG-based paths for drug repurposing, it is crucial to incorporate biomedical domain knowledge to guide the path-finding process. \citet{Liu_Yushan_2021} developed an RL-based model "PoLo" that utilizes the biological meta-paths identified in \citet{Himmelstein_Daniel_Scott_2017} via the "DWPC" method to supervise path searching for drug repurposing. However, the "PoLo" model does not scale to a massive and complex BKG (e.g., CKG and RTX-KG2) due to its dependence on the "DWPC" method that is reported to be computationally inefficient \citep{Womack_Finn_2019}. 

In this article, we describe \texttt{KGML-xDTD}: a \textbf{K}nowledge \textbf{G}raph-based \textbf{M}achine \textbf{L}earning framework for e\textbf{x}plainably predicting \textbf{D}rugs \textbf{T}reating \textbf{D}iseases, which contains two modules for both drug repurposing prediction and explanation. We propose to amplify the ability of RL model in biologically meaningful path searching by utilizing the biologically meaningful "demonstration paths" and pre-trained drug-repurposing model probability as rewards. We incorporate this idea into the appropriate models (e.g., GraphSAGE \citep{Hamilton_William_L_2017}, Random Forest, and ADAC RL \citep{Zhao_Kangzhi_2020} models) and then make them applicable to the explainable drug repurposing problem at massive data scale and complexity. By comparing with the existing popular drug repurposing models and evaluating the predicted paths with an expert-curated path-based drug MOA database \emph{DrugMechDB} \citep{Mayers_Michael_2020}, we show that the proposed model framework can achieve state-of-the-art performance in both predictions of drug repurposing and recapitulation of human-curated drug MOA paths provided by DrugMechDB. In further case studies, by comparing the model predictions with the real regulatory networks, we show that the proposed framework effectively identifies biologically reasonable BKG-based paths for real-world applications.

\section{Materials and Methods}\label{sec:materials_and_methods}
\subsection{Datasets}\label{subsec:datasets}

\subsubsection{Customized Biomedical Knowledge Graph}\label{subsubsec:cutomized_bkg}
\begin{figure*}
\centering
\includegraphics[width=1\textwidth]{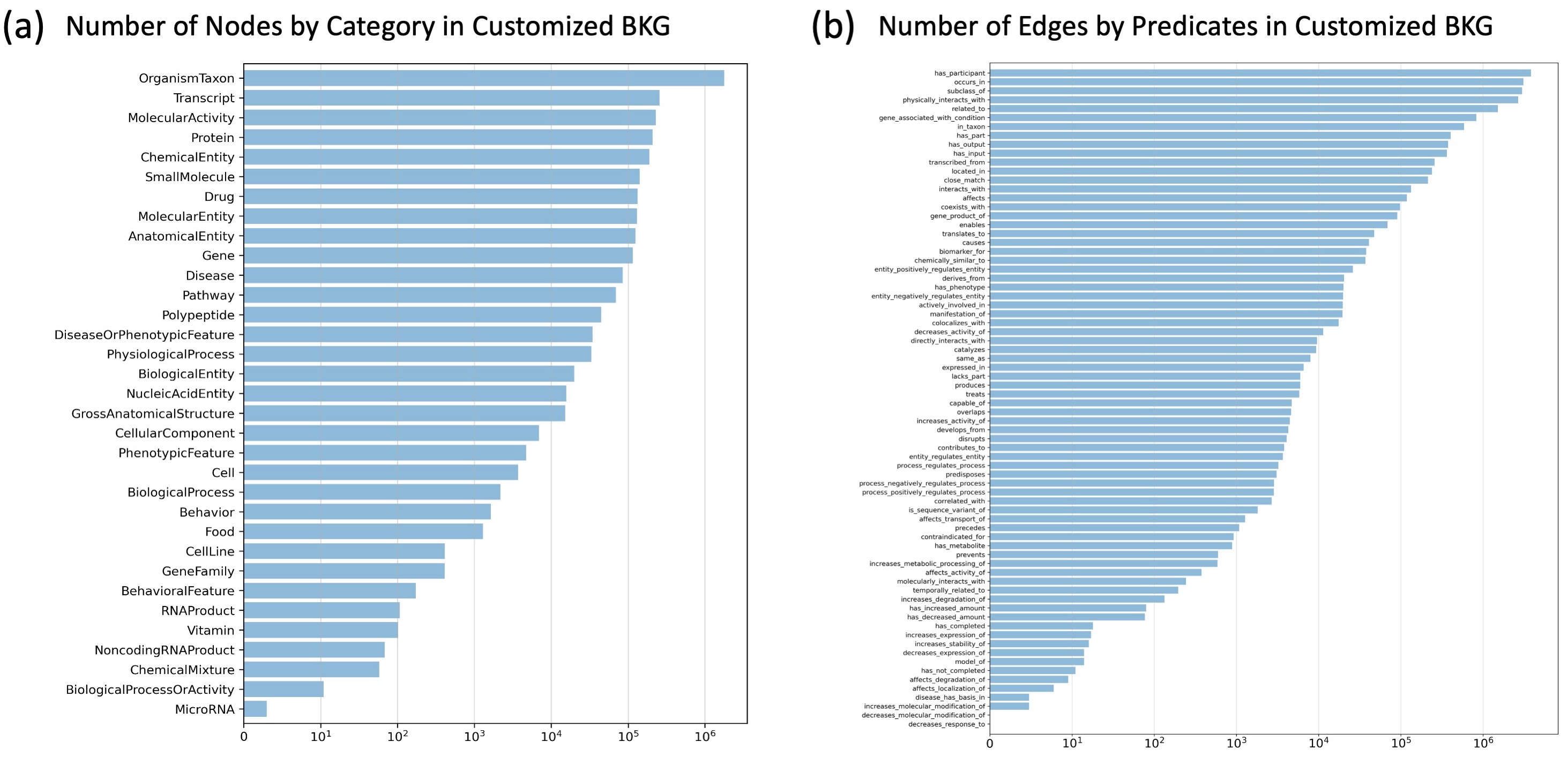}
\caption{Number of nodes by category (a) and number of edges by predicate (b) in customized Biomedical Knowledge Graph (BKG).}\label{fig:fig1}
\end{figure*}

To accommodate biomedical-reasonable predictions of drugs' indications and their mechanisms of action, the ideal biomedical knowledge graph should integrate biomedical knowledge from comprehensive and diverse databases and publications, as well as accurately identify and merge different identifiers representing the same biological entity into one (For example, "CHEBI:2367" and "CHEMBL455626" are two distinct identifiers separately presented in ChEBI database \citep{Degtyarenko_Kirill_2008} and ChEMBL database \citep{Gaulton_Anna_2012} but represent the same compound "abyssinone I"). Thus, we utilize the canonicalized version of the Reasoning Tool X Knowledge Graph 2 (\emph{RTX-KG2c}) \citep{Wood_E_C_2022}, one of the largest open-source biomedical knowledge graph (BKG) that integrates knowledge from extensive human-curated and Publication-based databases, and has been widely used in the Biomedical Data Translator Project \citep{Consortium_2019, Biomedical_Data_Translator_Consortium_2019}. Compared to other commonly used open-source BKGs mentioned above, \emph{RTX-KG2c} is a biolink-model-based \footnote{The biolink model \citep{unni2022biolink} is a universal and standardized BKG ontology framework} standardized \citep{unni2022biolink} and regularly-updated BKG that efficiently merges biologically and semantically equivalent nodes and edges via multiple curation steps. The version 2.7.3 of \emph{RTX-KG2c} that we use contains around 6.4M nodes and 39.3M edges with knowledge from 70 public biomedical sources, where all biological concepts (e.g., "ibuprofen") are represented as vertices and all concept-predicate-concept (e.g., "ibuprofen - increases activity of - GP1BA gene") are presented as edges. For drug repurposing purposes, we customized \emph{RTX-KG2c} with four principles (Please see more details in Appx.~\ref{sec:bkg_preprocessing}): 1). excluding the nodes whose categories are irrelevant to drug repurposing explanation (e.g., "GeographicLocation" and "Device"); 2). filtering out the low-quality edges based on our criteria; 3). removing the hierarchically redundant edges; 4). excluding all drug-disease edges. After these processing steps, 3,659,165 nodes with 33 distinct categories (Figure~\ref{fig:fig1} a) and 18,291,237 edges with 74 distinct types (Figure~\ref{fig:fig1} b) are left in our customized biomedical knowledge graph, which is used for downstream model training. 

\subsubsection{Data Sources for Model Training}\label{subsubsec:data_sources_for_model_training}
To train the \texttt{KGML-xDTD} framework for drug repurposing prediction and its explanation, we utilize four human-curated and NLP-derived training datasets:
\begin{itemize}
\item \textbf{MyChem Data} \citep{Xin_Jiwen_2018} is provided by the BioThings API collection \citep{Xin_Jiwen_2016}, which contains up-to-date human-curated annotations regarding indication and contraindication for drugs. We use drug-disease pairs with the relation "indication" as true positives while those with "contraindication" as true negatives.
\item \textbf{SemMedDB Data} \citep{Kilicoglu_Halil_2012} is provided by the Semantic MEDLINE Database (SemMedDB), which leverages natural language processing (NLP) techniques to extract semantic triples with "treats" and "negatively treats" relations from PubMed abstracts. We use drug-disease pairs with the relation "treats" as true positives while those with "negatively treats" as true negatives.
\item \textbf{NDF-RT Data} \citep{Brown_Steven_H_2004} is provided by National Drug File – Reference Terminology from Veterans Health Administration (VHA) which contains human-curated information on drug interaction, indications, and contraindications. We use drug-disease with therapeutics label "indications" as true positives while those with "contraindications" as true negatives.
\item \textbf{RepoDB Data} \citep{Brown_Adam_S_2017} is a standard set of successful and failed drug-disease pairs in clinical trials collected by the Blavatnik Institute at Harvard Medical School. We use drug-disease with the status "approved" as true positives while those with "withdrawn" as true negatives.
\end{itemize}
We further filter drug-disease pairs from SemMedDB Data due to publication bias and possible NLP mistakes by using both the co-occurrence frequency and the PubMed-publication-based Normalized Google Distance (NGD) \citep{Cilibrasi_R_L_2007} defined below: 
\begin{equation}\label{eq:eq1}
\footnotesize
NGD(c1, c2) = \frac{max\{log\mathcal{N}(c1), log\mathcal{N}(c2)\} - log\mathcal{N}(c1, c2)}{logN - min\{log\mathcal{N}(c1), log\mathcal{N}(c2)\}}
\end{equation}
where $c1$ and $c2$ are two biological concepts used in the customized BKG; $\mathcal{N}(c1)$ and $\mathcal{N}(c2)$ respectively represent the total number of unique PubMed IDs associated with $c1$ and $c2$; $\mathcal{N}(c1, c2)$ is the total number of unique PubMed IDs shared between $c1$ and $c2$; $N$ is the total number of pairs of Medical Subject Heading (MeSH) terms annotations in PubMed database.
Only the SemMedDB drug-disease pairs with at least 10 supporting publications and an NGD score of 0.6 or lower are left for the downstream model training. 

These datasets are pooled together and then processed by 1). mapping the raw identifiers of drugs and diseases to the identifiers used in the customized BKG; 2) removing duplicate drug-disease pairs in both the true positive set and the true negative set. Table~\ref{tab:source_table} shows the drug-disease pair count from each data source after data pre-processing.

\begin{table}[bth!]
	\caption{Pair count of true positive (indications) and true negative (contraindications or no effect) data from four data sources after data pre-processing.}\label{tab:source_table}
    \footnotesize
	\centering
    \begin{threeparttable}
	\begin{tabular}{lcc}
	    \toprule
		\textbf{Source} & \textbf{True Positive (Treats)} & \textbf{True Negative (Not Treat)} \\
		\midrule
		MyChem & 3,663  & 26,795  \\
		SemMedDB & 8,255 & 11    \\
		NDF-RT   & 3,421 & 5,119  \\
		RepoDB   & 2,127 & 738  \\
		Shared & 3,971 & 526 \\
		\midrule
		\textbf{Total} & 21,437 & 33,189 \\
		\bottomrule
	\end{tabular}
    \begin{tablenotes}
    \item Note that ‘shared’ means those pairs are from two or more data sources.
    \end{tablenotes}
    \end{threeparttable}
\end{table}

\subsubsection{DrugMechDB}\label{subsubsec:drugmechdb}
DrugMechDB\footnote{\url{https://sulab.github.io/DrugMechDB}} \citep{Mayers_Michael_2022}, to our best knowledge, is the first human-curated path-based database for explaining the mechanism of action (MOA) from a drug to a disease in an indication, with 3,593 MOA paths for 3,327 unique drug-disease pairs. These paths are extracted from free-text descriptions from DrugBank, Wikipedia, and other literature sources, and then have been curated by subject matter experts and also follow the schema of Biolink model. Hence, we can match them to nodes and edges used in the RTX-KG2 BKG via the \textit{Node Synonymizer} function \citep{Wood_E_C_2022}. Since the length of predicted MOA paths generated by the \texttt{KGML-xDTD} framework is fixed to 3 in this study, we consider those 3-hop BKG-based paths of which all four nodes show up in the complete DrugMechDB MOA paths are the correct matched paths. Thus, we find 472 unique drug-disease pairs, of which each has at least one such correct matched path in all possible 3-hop paths between drug and disease in the customized BKG. We use these matched paths as true positive biologically meaningful paths for the evaluation of the model-predicted paths. 

\subsection{Model Framework}\label{subsec:model_framework}
\begin{figure*}[h]
\centering
\begin{tikzpicture}
\draw (0,0 ) node[inner sep=0] {\includegraphics[width=1\textwidth, trim={1.7cm 6.55cm 0.7cm 0.92cm}, clip]{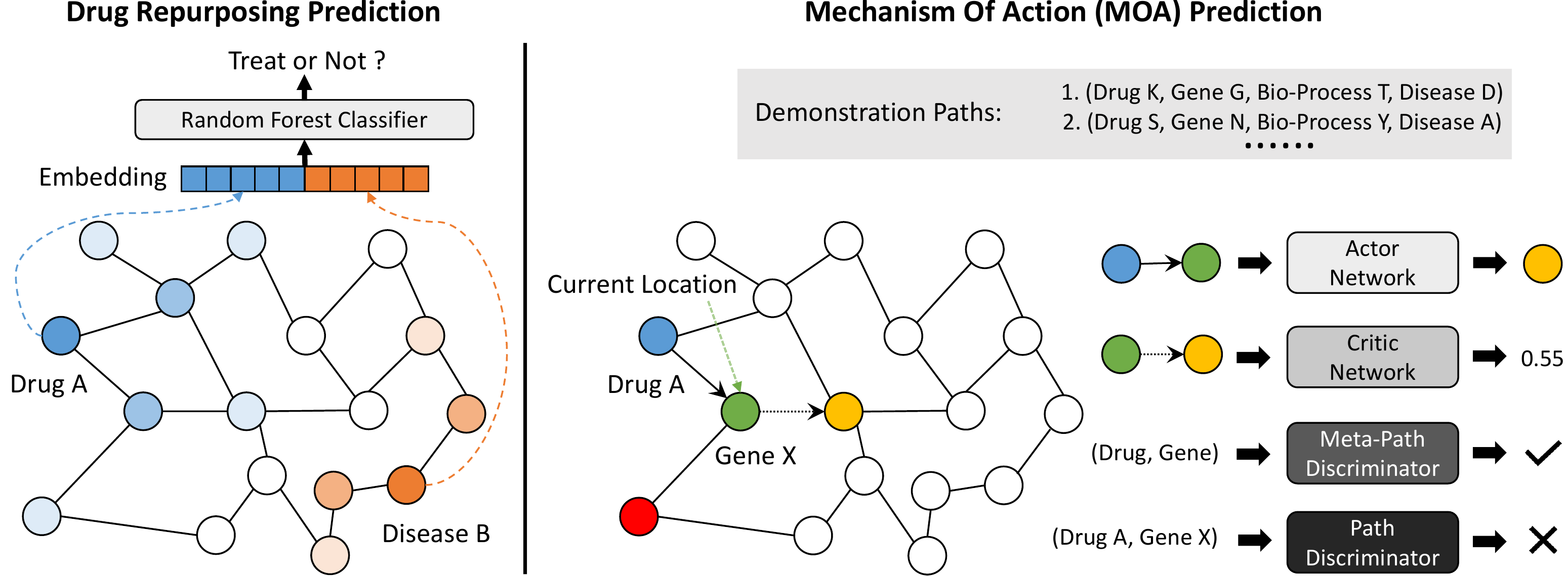}};
\end{tikzpicture}
\caption{Illustration of entire \texttt{KGML-xDTD} model framework: drug repurposing prediction module (left) and mechanism of action (MOA) prediction module (right).}\label{fig:fig2}
\end{figure*}

The model framework of \texttt{KGML-xDTD} consists of two modules: a drug repurposing prediction module that combines the advantages of GraphSAGE \citep{Hamilton_William_L_2017} and a Random Forest model, and an MOA prediction module that utilizes an adversarial actor-critic reinforcement learning (RL) model. We show the overview of the entire model framework in Figure~\ref{fig:fig2}. The implementation details of each module in \texttt{KGML-xDTD} framework are presented in Appx~\ref{sec:implemention_KGML_xDTD}.

\subsubsection{Notations}\label{subsubsec:notations}
Let $\mathcal{G} = \{\mathcal{V}, \mathcal{E}\}$ be a directed biomedical knowledge graph, where each node $v \in \mathcal{V}$ represents a biological entity (e.g., a specific drug, disease, gene, or pathway, etc.) and each edge $e \in \mathcal{E}$ represents a biomedical relationship (e.g., \textit{interacts-with, see more in Figure~\ref{fig:fig1} b}). We use $\mathcal{V}^{\mathrm{drug}}$ to represent all the drug nodes (the nodes with the categories of "Drug" and "Small Molecule" in the customized BKG) and $\mathcal{V}^{\mathrm{disease}}$ to represent all the disease nodes (the nodes with the categories of "Disease", "PhenotypicFeature", "BehavioralFeature" and "DiseaseOrPhenotypicFeature" in the customized BKG). For each notation, we use bold formatting to represent its embedding (e.g., $\bm{v}$ represents the embedding of $v$ ).

\subsubsection{Drug Repurposing Prediction}\label{subsubsec:drug_repurposing_prediction}
Drug repurposing aims to identify new indications of existing drugs/compounds. We solve it as a link prediction problem on the graph $\mathcal{G}$. Specifically, given any drug-disease pair $(v_i, v_j)$ where $v_i \in \mathcal{V}^{\mathrm{drug}}$ and $v_j \in \mathcal{V}^{\mathrm{disease}}$, we predict the probability that drug $i$ can be used to treat disease $j$.
We first use GraphSAGE to calculate the embedding for each node. Ideally, the node embeddings should contain two kinds of information: node attributes and node neighborhoods.

To capture the neighborhood information, we optimize GraphSAGE to encourage neighbor nodes to have similar embeddings and non-neighbor nodes to have distinct embeddings. Specifically, we perform random walks for each node to collect its neighborhood information and train the model to maximize a node's similarity with its neighbor nodes. For a node $u$, the loss is calculated as:
\begin{equation}
\footnotesize
L_{\mathcal{G}}(\bm{z}_u) = - \log (\sigma(\bm{z}_u^{\top}\bm{z}_v)) - k \cdot \mathbb{E}_{v_n \sim P_n(v)} \log (\sigma(\bm{z}_u^{\top}\bm{z}_{v_n}))
\end{equation}
where $\bm{z}_u, \bm{z}_v$ are respectively the embeddings of nodes $u,v$, $\sigma$ is the sigmoid function, $v$ is a node that co-occurs with $u$ in fixed-length random walks, $P_n$ represents negative sampling distribution, and $k$ indicates the number of negative samples (nodes not in $u$'s fixed-length neighborhood). 

To capture the node attributes information, we utilize the PubMedBERT model \citep{Gu_Yu_2022}, a pre-trained language model designed for biomedical texts, to generate a node attribute embedding for each node based on the concatenation of the node's name and category. We further compress the embeddings to 100 dimensions with Principal Components Analysis (PCA) to reduce memory usage and use them as the initial node feature for GraphSAGE. In this way, the final GraphSAGE embedding of each node should contain the information regarding both graph topology and node attributes. We concatenate the GraphSAGE embeddings of drug-disease pairs and use them as input of a Random Forest model to classify each drug-disease pair into one of the "not treat", "treat", and "unknown" classes. We obtain "treat" and "not treat" drug-disease pairs from four data sources (described in Sec.~\ref{subsubsec:data_sources_for_model_training}). We generate "unknown" drug-disease pairs through negative sampling \citep{Mikolov_Tomas_2013}, that is, replacing the drug or disease identifier in each "treat" drug-disease pair with a random drug or disease identifier to generate a new pair that does not appear in both the "treat" and "not treat" classes. Specifically, for each unique "treat" drug-disease pair, we respectively replace its drug identifier with 30 other random drug identifiers as well as replace its disease identifier with 30 other random disease identifiers to make 60 new drug-disease pairs for the "unknown" class.

\subsubsection{Mechanism of Action (MOA) Prediction}\label{subsec:moa_prediction}
When potential indications of a given drug are identified by the drug repurposing prediction module, a natural yet essential question is: can we biologically explain the predictions? We solve this by employing a reinforcement learning (RL) model to predict the BKG-based MOA paths, which are the paths on the knowledge graph from drug nodes to disease nodes. These BKG-based MOA paths can semantically describe an abstract biological process of how a drug treats a disease. 

\paragraph{\textbf{Demonstration paths}}
To encourage the RL agent to terminate the path searching at the expected diseases through a biologically reasonable path, we leverage so-called "demonstration paths", a set of biologically likely paths (e.g., drug1-gene1-protein3-disease1), that explains the underlying reasons for why a drug can treat a disease. We extract 396,705 demonstration paths from the customized BKG using the known drug-target interactions collected from two curated biomedical data sources: DrugBank (v5.1) and Molecular Data Provider (v1.2)\footnote{\url{https://github.com/NCATSTranslator/Translator-All/wiki/Molecular-Data-Provider}}, as well as the PubMed-publication-based Normalized Google Distance (NGD) (see Equation \ref{eq:eq1}). We show more details regarding demonstration path extraction in Appx.~\ref{sec:implemention_dpe}.

\paragraph{\textbf{Adversarial Actor-critic Reinforcement Learning}}\label{paragraph:ADAC}
We formulate the MOA prediction as a path-finding problem and adapt the Adversarial Actor-Critic Reinforcement Learning model \citep{Zhao_Kangzhi_2020} to solve it. Reinforcement learning is defined as a Markov Decision Process (MDP) which contains:

\textbf{States}: Each state $s_{t}$ at time $t$ is defined as $s_{t}=~\big(v_{drug}, v_{t}, (v_{t-1}, e_{t}), \dots, (v_{t-K}, e_{t-(K-1)}) \big)$ where $v_{drug} \in \mathcal{V}^{\mathrm{drug}}$ is a given starting drug node; $v_{t} \in \mathcal{V}$ represents the node where the agent locates at time $t$; the tuple $\big(v_{t-K}, e_{t-(K-1)}\big)$ represents the previous $K$th node and $(K-1)$th predicate. For the initial state $s_{0}$, the previous nodes and predicates are substituted by a special dummy node and predicate. We concatenate the embedding of all nodes and predicates of $s_{t}$ to get the state embedding $\bm{s}_{t}$, where the node embeddings are node attribute embeddings generated with the PubMedBERT model (see Sec.~\ref{subsubsec:drug_repurposing_prediction}) and the predicate embeddings employ one-hot vectors.

\textbf{Actions}: The action space $A_{t}$ of each node $v_{t}$ includes a self-loop action $a_{self}$ and the actions to reach its outgoing neighbors in the graph $\mathcal{G}$. Due to memory limitation and extremely large outdegree of certain nodes in the knowledge graph, we prune the neighbor actions based on the PageRank scores if a node has more than 3,000 neighbors. Specifically, we let $A_{t}= (a_{self}, a_{1},\dots, a_{k},\dots, a_{n_{v_t}})$ where $n_{v_t}$ is out-degree of node $v_t \in \mathcal{V}$. For each action $a_{t}=(e_{t},v_{t+1}) \in A_{t}$ taken at time $t$, we concatenate its node and predicate embeddings to obtain action embedding $\bm{a}_t$. We learn two embedding matrices $E^{N_\mathrm{n} \times \mathrm{d}}$ and $E^{N_\mathrm{p} \times \mathrm{d}}$ respectively for nodes and predicates\footnote{Each sub-network uses separate embedding matrices.}, where $d$ represents the embedding dimension, $N_\mathrm{n}$ represents the number of nodes in graph, and $N_\mathrm{p}$ represents the number of predicate categories in graph.

\textbf{Rewards}: During the path searching process, the agent only receives a terminal reward $R_{e,T}$ from environment (that is, there is no intermediate reward from environment: $R_{e,t}=0, \forall t<T$). Let $v_T$ be the last node of the path, and $\mathcal{N}_{drug}$ be the known diseases that drug $v_{drug}$ can treat. The terminal reward $R_{e,T}$ from environment is calculated with the drug repurposing model via:
$$
\footnotesize
  R_{e,T}=\left\{
    \begin{array}{ll}
      1, & \mbox{if $v_T \in \mathcal{N}_{drug}$}.\\
      p_{treat}, & \mbox{if $v_{T} \notin \mathcal{N}_{drug} ; v_{T} \in \mathcal{V}^{\mathrm{disease}}$ and $f(v_{drug}, v_{T})$ is predicted as "treat"}.\\
      0, & \mbox{if $v_{T} \notin \mathcal{N}_{drug} ; v_{T} \in \mathcal{V}^{\mathrm{disease}}$ and $f(v_{drug}, v_{T})$ is not predicted as "treat"}.\\
      -1, & \mbox{if $v_{T} \notin \mathcal{V}^{\mathrm{disease}}$}.
    \end{array}
  \right.
$$
where $p_{treat}$ is the "treat" class probability predicted by the drug repurposing model $f$.

The Adversarial Actor-critic RL model consists of four sub-networks that share the same model architecture $\mathrm{MLP}^{i}$ (note that $i$ represents the id of each sub-network described later, such as $a$ for \textit{actor network}, $c$ for \textit{critic network}, etc.) but with different parameters:
\begin{equation}
\footnotesize
\mathrm{MLP}^{i}(X) = BA(BA(XW^i_{1}+b^i_{1})W^i_{2}+b^i_{2})W^i_{3}+b^i_{3}
\end{equation}
where \{${W^i_{1}, W^i_{2}, W^i_{3}, b^i_{1}, b^i_{2}, b^i_{3}}$\} are the parameters and biases of linear transformations, $BA$ represents a batch normalization layer followed by an ELU activation function.

\textbf{Actor network}: The actor network learns a path-finding policy $\pi_{\theta}$ (note that $\theta$ represents all parameters of actor network) to guide the agent to choose an action $a_{t}$ from the action space $A_{t}$ based on the current state $s_{t}$:
\begin{equation}
\footnotesize
\pi_{\theta}(a_{t}|s_{t}, A_{t}) = \mathrm{softmax}(\bm{A}_{t} \odot \mathrm{MLP}^{a}(\bm{s}_t))
\end{equation}
where $\bm{A}_{t}$ is the embedding matrix of the action space $A_{t}$; $\odot$ represents the dot product. Here, $\pi_{\theta}(a_{t}|s_{t}, A_{t})$ represents the probability of choosing action $a_{t}$ at time $t$ from the action space $A_{t}$ given the state $s_{t}$.

\textbf{Critic network}: The critic network \citep{Lillicrap_Timothy_P_2015} estimates the expected reward $Q_{\phi}(s_{t},a_{t})$ (note that $\phi$ represents all parameters of the critic network) if the agent takes the action $a_{t}$ at the state $s_{t}$ by:
\begin{equation}
\footnotesize
Q_{\phi}(s_{t},a_{t}) =  \mathrm{MLP}^{c}(\bm{s}_t) \odot \bm{a}_{t}
\end{equation}

\textbf{Path discriminator network}: Since the RL agent only receives a terminal reward $R_{e,T}$ from environment indicating whether it reaches an expected target, to encourage the agent to find biologically reasonable paths and provide intermediate rewards, we further guide it with demonstration paths.
This network is essentially a binary classifier that distinguishes whether a path segment $(s_{t}, a_{t})$ is from demonstration paths or generated by the actor network. We treat all the known demonstration path segments $(s_{t}^{D}, a_{t}^{D})$ as positive samples and all actor-generated non-demonstration path segments $(s_{t}^{ND}, a_{t}^{ND})$ as negative samples. The path discriminator network $D_{p}(s,a) = \mathrm{sigmoid}(\mathrm{MLP}^{p}(\bm{s} \oplus \bm{a}))$, where $\bm{s}$ and $\bm{a}$ are respectively the embeddings of the state $s$ and the action $a$; $\oplus$ represents the concatenation operator, is optimized with: 
\begin{equation}
\footnotesize
L_{p} = - \mathbb{E}_{(s,a) \sim P_{D}}[\log(D_{p}(s,a))] - \mathbb{E}_{(s,a) \sim P_{A}}[\log(1-D_{p}(s,a))]
\end{equation}
where $P_{D}$ and $P_{A}$ respectively represent the demonstration path segment distribution and the actor-generated non-demonstration path segment distribution. Based on the probability $D_{p}(s_{t},a_{t})$, the path-discriminator-based intermediate reward $R_{p,t}$ is calculated as:
\begin{equation}
\footnotesize
R_{p,t} = \log(D_{p}(s_{t},a_{t})) - \log(1-D_{p}(s_{t},a_{t})).
\end{equation}

\textbf{Meta-Path discriminator network}: Similar to the path discriminator, this network aims to judge whether the meta-path of the actor-generated paths is similar to that of demonstration paths. The meta-path is the path of node categories (e.g., ["Drug"→"Gene"→"BiologicalProcess"→"Disease"]). Similarly, the meta-path discriminator $D_{m}(M) = \mathrm{sigmoid}(\mathrm{MLP}^{m}(\bm{M}))$, where $\bm{M}$ is the embedding of the meta-path $M$ defined as the concatenation of learned category embeddings of all nodes that appear in the path, is also a binary classifier where the meta-paths of demonstration paths are treated as positive samples while others are negative samples. We optimize it with the following loss:
\begin{equation}
\footnotesize
L_{m}= - \mathbb{E}_{M \sim P^M_{D}}[\log(D_{m}(M))] - \mathbb{E}_{M \sim P^M_{A}}[\log(1-D_{m}(M))]
\end{equation}
where $P^M_{D}$ and $P^M_{A}$ respectively represent the demonstration meta-path distribution and the actor-generated non-demonstration meta-path distribution. The intermediate reward $R_{m,t}$ generated by the meta-path discriminator is calculated by:
\begin{equation}
\label{func:reward}
\footnotesize
R_{m,t} = \log(D_{m}(M)) - \log(1-D_{m}(M)). 
\end{equation}

The integrated intermediate reward $R_{t}$ at time $t$ is then calculated as:
\begin{equation}
\footnotesize
R_{t} = {\alpha}_{p}R_{p,t}+{\alpha}_{m}R_{m,t}+(1-{\alpha}_{p}-{\alpha}_{m}){\gamma}^{T-t}R_{e,T}
\end{equation}
where ${\alpha}_{p} \in [0,1]$ and ${\alpha}_{m} \in [0, 1-{\alpha}_{p}]$ are hyperparameters,  $\gamma$ is the decay coefficient, and $R_{e,T}$ is defined in the "Rewards" section above. 

To optimize the critic network, we minimize the Temporal Difference (TD) error \citep{Sutton_Richard_S_1988} with loss:
\begin{equation}
\footnotesize
L_{c}= \mathrm{TD}^{2} = [(R_{t} + Q_{\phi}(s_{t+1},a_{t+1})) - Q_{\phi}(s_{t},a_{t})]^{2}.
\end{equation}

Since the goal of the actor network is to achieve the largest expected reward by learning an optimal actor policy, we optimize the actor network by maximizing $J({\theta}) = \mathbb{E}_{a \sim {\pi}_{\theta}}[Q_{\phi}(s_{t},a)]$. We use the REINFORCE algorithm \citep{Williams_Ronald_J_1992} to optimize the parameters. To encourage more diverse exploration in finding paths, we use the entropy of ${\pi}_{\theta}$ as a regularization term and optimize the actor network with the following stochastic gradient of the loss function $L_a$:
\begin{equation}
\footnotesize
{\nabla}_{\theta}L_{a} = - {\nabla}_{\theta}J({\theta})= - \mathbb{E}_{{\pi}_{\theta}}[{\nabla}_{\theta}\mathrm{TD} \log{\pi}_{\theta}(a_{t}|s_{t})] - {\alpha}{\nabla}_{\theta}\mathrm{entropy}({\pi}_{\theta})
\end{equation}
where ${\pi}_{\theta}$ is the action probability distribution based on the actor policy, and $\alpha$ is the entropy weight.

We follow \citet{Zhao_Kangzhi_2020} to train the Adversarial Actor-critic RL model in a multi-stage way. First, we initialized the actor network using the behavior cloning method \citep{A_PomerleauDean_1991} in which the training set of demonstration paths is used to guide the sampling of the agent with Mean Square Error (MSE) loss. Then, in the first $z$ epochs, we freeze the parameters of the actor network and the critic network and respectively train the path discriminator network and meta-path discriminator network by minimizing $L_{p}$ and $L_{m}$. After $z$ epochs, we unfreeze the actor network and the critic network and optimize them together by minimizing a joint loss $L_{joint} = L_{a} + L_{c}$.

\section{Results}\label{sec:results}

\subsection{Evaluation Settings}\label{subsec:evaluation_settings}

\subsubsection{Data Split}\label{subsubsec:data_split}
The post-processed drug-disease pairs (described in Sec.~\ref{subsubsec:data_sources_for_model_training}) are split into training, validation, and test sets where the drug-disease pairs of each unique drug are randomly split according to a ratio of 8/1/1. For example, let’s say drugA has 10 known diseases that it treats (e.g., drugA-disease1, …, drugA-disease10), 8 pairs are randomly split into the training set, 1 pair is to the validation set, 1 pair to the test set. With this data split method, the model can be exposed to every drug in the training set, which complies with our goal of predicting new indications of known drugs and their potential mechanisms of action (MOAs) based on the MOA of known target diseases.
 
\subsubsection{Evaluation Metrics}\label{subsubsec:evaluation_metrics}
The proposed framework \texttt{KGML-xDTD} is evaluated on two types of tasks: predicting drug-disease "treat" probability (i.e., drug repurposing prediction) as well as identifying biologically reasonable BKG-based MOA paths from all candidates (i.e., MOA prediction). These two tasks are evaluated based on classification accuracy-based metrics (e.g., accuracy, macro f1 score) and ranking-based metrics (e.g., mean percentile rank, mean reciprocal rank, and proportion of ranks smaller than K) defined as follows:

\textbf{Accuracy (ACC)} is the fraction of the model classification is correct, computed as:  
\begin{equation}
\footnotesize
\mbox{ACC} = \frac{\mbox{Number of correct classifications}}{\mbox{Total number of drug-disease pair classifications}}
\end{equation}

\textbf{Macro F1 score (Macro-F1)} is the unweighted mean of all the per-class F1 scores:
\begin{equation}
\footnotesize
F1^{c} = 2*\frac{precision^{c} \times recall^{c}}{precision^{c} + recall^{c}}
\;\;\;\;\;\;\;\;
\mbox{Macro-F1} = \frac{1}{|C|} \sum_{c \in C}{F1^{c}}
\end{equation}
where $C$ presents classification classes (e.g.,"treat," "not treat," and "unknown").

\textbf{Mean Percentile Rank (MPR)} is the average percentile rank of the 3-hop correct DrugMeshDB-based matched path (described in Sec.~\ref{subsubsec:drugmechdb}) of true positive drug-disease pairs: 
\begin{equation}
\footnotesize
\mbox{MPR} = \frac{1}{|PR|} \sum_{pr \in PR}{pr}
\end{equation}
where $PR$ is a list of percentile ranks of correct matched paths of true positive drug-disease pairs ("treat" category).

\textbf{Mean Reciprocal Rank (MRR)} is the average inverse rank of true positive drug-disease pairs ("treat" category) or their 3-hop correct matched paths:
\begin{equation}
\footnotesize
\mbox{MRR} = \frac{1}{|R|} \sum_{r \in R}{\frac{1}{r}}
\end{equation}
where $R$ is a list of ranks of true positive drug-disease pairs or their correct matched paths.

\textbf{Hit@K} is the proportion of ranks smaller thank K for true positive drug-disease pairs ("treat" category) or their 3-hop correct matched paths:
\begin{equation}
\footnotesize
\mbox{Hit@K} = \frac{1}{|R|} \sum_{r \in R}{|r \leq k|}
\end{equation}
where $R$ is a list of ranks of true positive drug-disease pairs or their correct matched paths.

\subsubsection{Drug Repurposing Prediction Evaluation Method}\label{subsubsec:drug_repurposing_prediction_evaluation_method}
We utilize the metrics \textit{ACC} and \textit{Macro-F1} to measure the accuracy of drug repurposing prediction of our \texttt{KGML-xDTD} framework while using ranking-based metrics \textit{MRR} and \textit{Hit@K} to show its capability in reducing false positive (i.e., the false drug-disease pairs ranking higher among possible drug-disease candidates). We use the following three methods to generate non-true-positive drug-disease candidates for each true positive drug-disease pair for the \textit{MRR} and \textit{Hit@K} calculation:
\begin{itemize}
    \item \textbf{Drug-rank-based replacement}: For each true positive drug-disease pair, the drug-rank-based replacement pairs are generated by replacing the drug entity with each of all 274,676 other drugs in the customized BKG while excluding all known true positive drug-disease pairs.
    \item \textbf{Disease-rank-based replacement}: For each true positive drug-disease pair, the disease-rank-based replacement pairs are generated by replacing the disease entity with each of all 124,638 other diseases in the BKG while excluding all known true positive drug-disease pairs.
    \item \textbf{Combined Replacement}: For each true positive drug-disease pair, the combined replacement pairs are the combination of all replacement pairs of the above two methods. All known true positive drug-disease pairs are excluded from these replacement pairs.
\end{itemize}

Due to the massive size of possible drug-disease candidates, some baseline models (e.g., GAT and GraphSAGE+SVM) are not applicable in this setting within a reasonable time (e.g., a week). Thus, we also use a small subset of drug-disease replacement (we use 1000 random drug-disease pairs from the combined replacement set above: 500 with drug id replacement and 500 with disease id replacement) to calculate \textit{MRR} and \textit{Hit@K} in order to compare \texttt{KGML-xDTD} with all baselines. In addition, since the drug repurposing prediction module of \texttt{KGML-xDTD} framework does 3-class classification while other baselines do 2-class classification, for a fair comparison, we re-calculate \textit{ACC} and \textit{Macro-F1} for \texttt{KGML-xDTD} by excluding the "unknown" class.

\subsubsection{MOA Prediction Evaluation Method}\label{subsubsec:moa_prediction_evaluation_method}
For the evaluation of MOA prediction, we use the DrugMechDB \citep{Mayers_Michael_2020} to obtain the expert-verified MOA paths as ground-truth data and match each biological concept in these verified MOA paths to the biological entities used in the customized BKG, and then generate the correct matched paths (described in Sec.~\ref{subsubsec:drugmechdb}). we first calculate the path scores for all 3-hop KG paths between drug and disease with the path-finding policy learned from Adversarial Actor-critic Reinforcement Learning (RL) model using equation: 
\begin{equation}
\footnotesize
\mbox{path score} = \sum_{i=1}^{k}{{\delta}^{i-1} \times \log(P_{i} \times N_{i})}
\end{equation}
where $k$ is the number of hops in this path; $\delta$ is a decay coefficient (we set it to 0.9 in this study); $P_{i}$ represents the probability of choosing action $a_{i}$ in the $i^{\rm th}$ hop following this path based on the trained RL model; $N_{i}$ is the number of possible actions in the $i^{\rm th}$ hop. 

With these path scores, we obtain the ranks of the matched BKG-based MOA paths and calculate the ranking-based metrics \textit{MPR}, \textit{MRR}, and \textit{Hit@K}. For those drug-disease pairs with multiple BKG-based MOA paths, we use the highest ranks of their paths. We compare \texttt{KGML-xDTD} with the baseline models based on these metrics to show the capability of MOA prediction module of \texttt{KGML-xDTD} in identifying biologically reasonable BKG-based MOA paths from a massive and complex BKG with comparably low false positive. In addition, we further perform two case studies to evaluate the effectiveness of \texttt{KGML-xDTD} in identifying the biologically reasonable BKG-based paths.

\subsection{Drug Repurposing Prediction Evaluation}\label{subsec:drug_repurposing_prediction_evaluation}

For drug repurposing prediction evaluation, we compare the \texttt{KGML-xDTD} model framework against several state-of-the-art (SOTA) KG-based models and variants of \texttt{KGML-xDTD} for drug repurposing prediction based on the method described in Sec. "Drug Repurposing Prediction Evaluation Method" on pag~\ref{subsubsec:moa_prediction_evaluation_method}.

We use eight different SOTA KG-based models as baseline models that are commonly used for BKG-based drug repurposing \citep{Zhang_Rui_2021, Hsieh_Kanglin_2020}. \texttt{TransE} \citep{Bordes_Antoine_2013},  \texttt{TransR} \citep{Yankai_Lin_2015},  \texttt{RotatE} \citep{Sun_Zhiqing_2019} are the translation-distance-based models that regard a relation (e.g., "treats") as a "translation"/"rotation" (e.g., a kind of spatial transformation) from a head entity (e.g., a drug node) to a tail entity (e.g., a disease node).  \texttt{DistMult} \citep{Yang_Bishan_2014} is a bilinear model that measures the latent semantic similarity of a knowledge-graph triple (head entity, relation/predicate, tail entity) with a trilinear dot product. \texttt{ComplEx} \citep{Trouillon_Theo_2016} and \texttt{ANALOGY} \citep{Hanxiao_Liu_2017} are the extensions of \texttt{DistMult} that consider more complex relations (e.g., asymmetric relations). \texttt{SimpLE} \citep{Kazemi_Seyed_Mehran_2018} is a tensor-factorization-based model to learn the semantic relation of a knowledge-graph triple. \texttt{GAT} \citep{velickovic_2018} is a popular graph neural model that leverages the important graph topology structure based on self-attention mechanism for graph-associated tasks (e.g., link prediction). Implementation details of these baselines are presented in Appx.~\ref{sec:implemention_baselines}.

Besides these SOTA baseline models, we also compare the drug repurposing prediction module in \texttt{KGML-xDTD} with its several variants to show the effectiveness of model components. For example, to show efficacy of the combination of GraphSage and Random Forest (RF), we use a pure GraphSAGE for link prediction (\texttt{GraphSAGE-link}), the combination of GraphSage and logistic model (\texttt{GraphSAGE-logistic}), and the combination of GraphSage and Support Vector Machine (SVM) model (\texttt{GraphSAGE-SVM}). To demonstrate the effectiveness of node attribute embeddings (described in Sec.~\ref{subsubsec:drug_repurposing_prediction}) in improving repurposing prediction, we conduct an ablation experiment that replaces node attribute embeddings (NAEs) with random embeddings (initialized with the Xavier method \citep{Glorot_Xavier_2010}) as GraphSage initialized embeddings (\texttt{KGML-xDTD w/o NAE}); to support rationality of setting "unknown" class through negative sampling  (described in Sec.~\ref{subsubsec:drug_repurposing_prediction}), we modify the drug repurposing prediction module for 2-class classification\footnote{only considers true positive and true negative.} (\texttt{2-class KGML-xDTD}) as a baseline comparison model.

\begin{table*}[bth!]
    \caption{The performance comparison of Drug Repurposing Prediction (DRP) between \texttt{KGML-xDTD} and different baseline models based on test dataset (described in Sec.~\ref{subsubsec:data_split}). The top panel shows the performance of state-of-the-art (SOTA) baseline models; the middle panel shows the performance of variants of \texttt{KGML-xDTD} model framework; the bottom panel shows the performance of \texttt{KGML-xDTD} model framework.}\label{tab:drug_prepurposing_evaluation_table}. 
    \footnotesize
	\centering
    \begin{threeparttable}
	\begin{tabular}{lcccccc}
	    \toprule
		\textbf{Model} & \textbf{Accuracy} & \textbf{Macro F1 score} & \textbf{MRR} & \textbf{Hit@1} & \textbf{Hit@3} & \textbf{Hit@5} \\
		\midrule
		TransE & 0.708 & 0.708 & 0.280 & 0.117 & 0.301 & 0.449 \\
		TransR & 0.858 & 0.855 & 0.301 & 0.123 & 0.344 & 0.518 \\
		RotatE & 0.704 & 0.704 & 0.253 & 0.075 & 0.283 & 0.450 \\
		DistMult & 0.555 & 0.495 & 0.173 & 0.041 & 0.143 & 0.259  \\
		ComplEx & 0.624 & 0.460 & 0.132 & 0.020 & 0.106 & 0.194 \\
		ANALOGY & 0.594 & 0.465 & 0.180 & 0.045 & 0.147 & 0.274  \\
		SimplE & 0.599 & 0.472 & 0.163 & 0.038 & 0.137 & 0.244 \\
		GAT & \textbf{0.936} & \textbf{0.934} & 0.003 & 0.001 & 0.001 & 0.001  \\
        \midrule
		GraphSAGE-link & 0.919 & 0.915 & 0.002 & 0 & 0 & 0  \\
	    GraphSAGE+logistic & 0.791 & 0.784 & 0.002 & 0 & 0 & 0  \\
	    GraphSAGE+SVM & 0.807 & 0.793 & 0.002 & 0 & 0 & 0 \\
        KGML-xDTD w/o NAEs & 0.909 (0.898$^{\ast}$) & 0.891 (0.892$^{\ast}$) & 0.150 & 0.029 & 0.138 & 0.243 \\
	    2-class KGML-xDTD & 0.929 & 0.925 & 0.271 & 0.177 & 0.310 & 0.381   \\
		\midrule
		KGML-xDTD (ours) & 0.935 (0.930$^{\ast}$) & 0.923 (0.926$^{\ast}$)  & \textbf{0.356} & \textbf{0.206} & \textbf{0.407} & \textbf{0.522}  \\
		\bottomrule
	\end{tabular}
     \begin{tablenotes}
    \item 1. The values with $^{\ast}$ inside the parenthesis are the adjusted results by excluding the "unknown" category for a fair comparison.
    \item 2. The ranking metrics (e.g., "MRR" and "Hit@K") are calculated with the random drug-disease replacement method (i.e., we use 1,000 random drug-disease pairs to calculate the ranks of true positive drug-disease pairs in test dataset).
    \item 3. "NAEs" represents node attribute embeddings.
    \end{tablenotes}    
    \end{threeparttable}
\end{table*}

\begin{figure*}[h]
\centering
\includegraphics[width=1\textwidth]{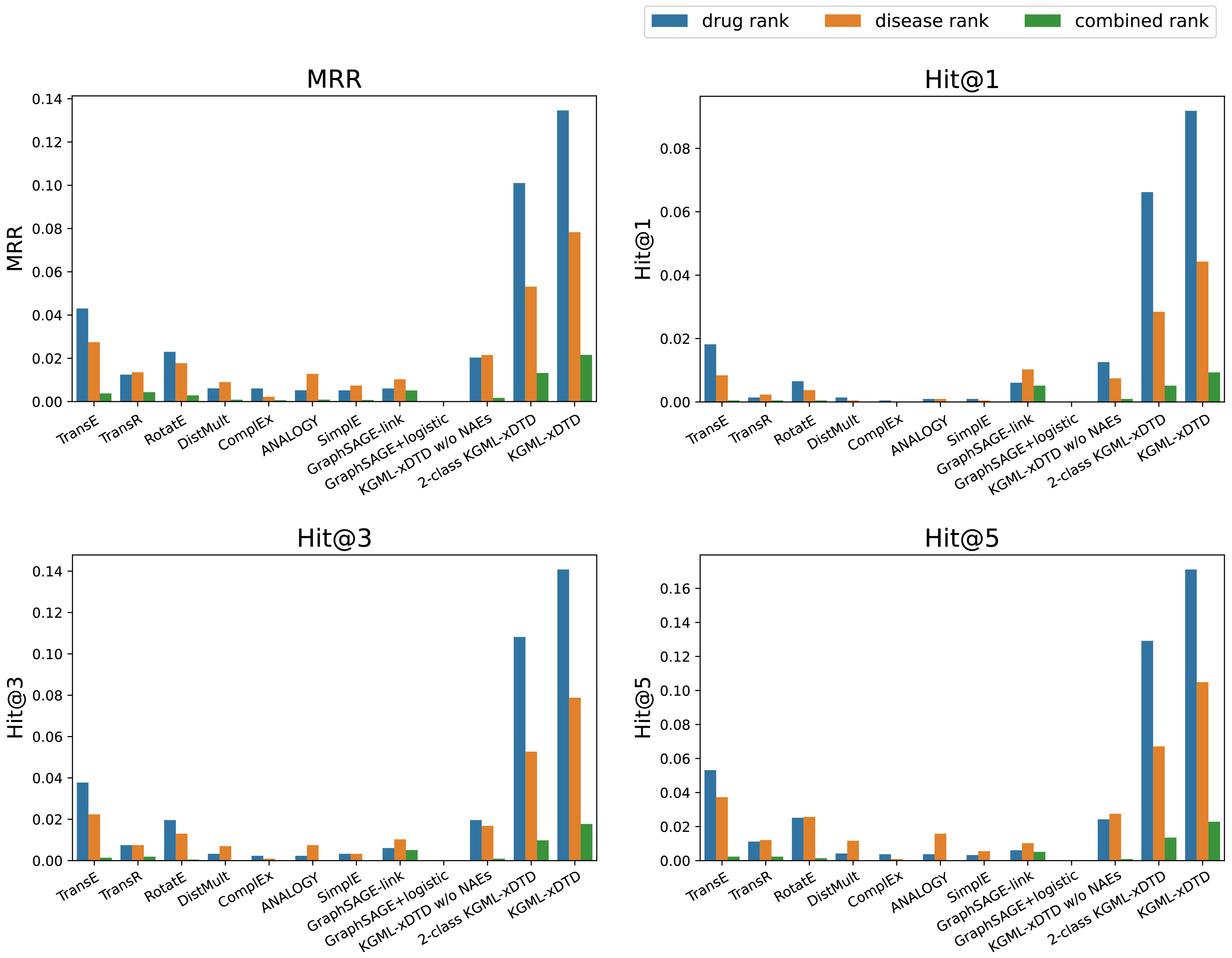}
\caption{The performance comparison of Drug Repurposing Prediction (DRP) between \texttt{KGML-xDTD} and different baseline models (\texttt{GAT} and \texttt{GraphSAGE+SVM} are excluded due to computation time constraints) based on test dataset using three "complete" replacement  methods (i.e., Drug-rank-based replacement", "Disease-rank-based replacement", and "Combined replacement") to generate non-true-positive drug-disease candidates for each true positive drug-disease pair for MRR and Hit@K calculation (see more details in Sec.~\ref{subsubsec:drug_repurposing_prediction_evaluation_method}). The legend "drug rank", "disease rank" and "combined rank" respectively correspond to the methods of "Drug-rank-based replacement", "Disease-rank-based replacement", and "Combined replacement" described in Sec.~\ref{subsubsec:drug_repurposing_prediction_evaluation_method}.}\label{fig:fig3}
\end{figure*}

Table~\ref{tab:drug_prepurposing_evaluation_table} shows the performance of \texttt{KGML-xDTD} model framework and other baseline models in the task of drug repurposing prediction based on the metrics described in Sec.~\ref{subsubsec:evaluation_metrics} and test dataset along with 1,000 random drug-disease replacement pairs. As shown in the table, on the one hand, the \texttt{KGML-xDTD} outperforms most of the baseline models and achieves comparable performance as \texttt{GAT} in classification-based metrics (e.g., accuracy, macro f1 score), indicating its effectiveness in classifying known "treat" and "not treat" drug-disease pairs with both attribute and neighborhood information on the knowledge graph. On the other hand, \texttt{KGML-xDTD}'s exceptional performance in ranking-based metrics shows its superiority over baselines in identifying new indications of existing drugs out of a large number of possible drug-disease pairs with relatively low false positives (Figure~\ref{fig:fig3} further supports this conclusion with three different "complete" replacement methods), which is of great importance for guiding clinical research. Besides, by comparing \texttt{2-class KGML-xDTD} with the vanilla GraphSAGE model (e.g., \texttt{GraphSAGE-link}), we demonstrate the effectiveness of the Random Forest model over a neural network classifier in this task. The comparison between \texttt{KGML-xDTD w/o NAE} and \texttt{KGML-xDTD} shows that the \texttt{KGML-xDTD} benefits from the use of node attribute embeddings for drug repurposing prediction while the comparison with \texttt{2-class KGML-xDTD} indicates the effectiveness of using negative sampling to generate "unknown" drug-disease pairs for model training. With the "unknown"  drug-disease pairs, the \texttt{KGML-xDTD} model achieves significant improvement in ranking-based metrics, which is essential when applying to real-world drug repurposing because it can reduce the false positives.

\subsection{MOA prediction evaluation}\label{subsec:moa_prediction_evaluation}
For MOA prediction, we evaluate how well the \texttt{KGML-xDTD} can identify the DrugMechDB-matched BKG-based MOA paths (described in Sec.~\ref{subsubsec:moa_prediction_evaluation_method}) from a large number of possible paths in the customized BKG by utilizing ranking-based metrics (e.g., MPR, MRR, and Hit@K) and two specific case studies.

There are few machine learning models designed for the task of identifying biologically meaningful paths from biomedical knowledge graphs for explaining drug repurposing. To our best knowledge, although three models (e.g., UKGE \citep{Sosa_Daniel_N_2020}, GrEDeL \citep{Sang_Shengtian_2019}, Polo \citep{Liu_Yushan_2021}, all mentioned in Sec.~\ref{sec:introduction}) were proposed and can be used for this goal, they all have certain constraints and cannot be used as baseline models for comparison. The UKGE model cannot be applied to BKGs without weighted edge information (e.g., frequency of relation appeared in literature). The authors of the GrEDeL model don't provide the code to implement this model. The Polo model cannot be trained within a reasonable time (e.g., within two weeks) on a massive and complex BKG (e.g., RTX-KG2) due to its dependence on a computationally inefficient method "DWPC" \citep{Himmelstein_Daniel_Scott_2017}. Therefore, we choose the \texttt{MultiHop} reinforcement learning model \citep{Lin_Xi_Victoria_2018} as a baseline model since it uses the LSTM model framework as the GrEDel model and allows using a self-defined reward shaping strategy in its reward function as what we do in the \texttt{KGML-xDTD} model (i.e., we can use the same reward strategy described in Sec.~\ref{paragraph:ADAC}). Furthermore, we also compare with an ablated version of \texttt{KGML-xDTD} (i.e., \texttt{KGML-xDTD w/o DP} which does not take advantage of the demonstration paths by setting $\alpha_p$ and $\alpha_m$ in Function \ref{func:reward} as $0$) as another baseline model to show the importance of demonstration paths.

\begin{table*}[bth!]
\centering
\caption{The performance comparison of Mechanism of Action (MOA) Prediction between \texttt{KGML-xDTD} and different baseline models (e.g., \texttt{MultiHop} and \texttt{KGML-xDTD w/o DP}) based on test dataset (described in Sec.~\ref{subsubsec:data_split}). The metrics in this table are calculated using path scores and all non-DrugMechDB-matched 3-hop paths between drug and disease as "negative" paths for each true positive drug-disease pair (see more details in Sec.~\ref{subsubsec:moa_prediction_evaluation_method}).}\label{tab:moa_prediction_evaluation_table}
\footnotesize
{
	\begin{tabular}{lccccccc}
	    \toprule
		\textbf{Model} & \textbf{MPR} & \textbf{MRR} & \textbf{Hit@1} & \textbf{Hit@10} & \textbf{Hit@50} & \textbf{Hit@100} & \textbf{Hit@500} \\
		\midrule
		MultiHop & 61.400\% & 0.027 & 0.017 & 0.042 & 0.067 & 0.118 & 0.345 \\
		KGML-xDTD w/o DP & 72.965\% & 0.015 & 0.008 & 0.017 & 0.067 & 0.160 & 0.403  \\
		\midrule
		KGML-xDTD (ours) & \textbf{94.696\%} & \textbf{0.109} & \textbf{0.059} & \textbf{0.193} & \textbf{0.496} & \textbf{0.613} & \textbf{0.849} \\
		\bottomrule
	\end{tabular}
}
\end{table*}

We show the comparison results between the \texttt{KGML-xDTD} model framework and different baseline models for MOA prediction evaluation in Table~\ref{tab:moa_prediction_evaluation_table}. Although all the models receive the same terminal rewards from the environment, that is given by the drug repurposing prediction module of \texttt{KGML-xDTD}, the MOA prediction module of \texttt{KGML-xDTD} achieves significantly better performance in identifying DrugMechDB-matched BKG-based MOA paths than the other two baselines across all ranking-based metrics. Comparison between the \texttt{KGML-xDTD} with and without demonstration paths (i.e., \texttt{KGML-xDTD w/o DP}) further illustrates the great effectiveness of using demonstration paths to guide the path-finding process. Due to the massive searching space and sparse rewards, the RL agent often fails to find reasonable paths out of many possible choices, while our model \texttt{KGML-xDTD}, with the intermediate guidance provided by the demonstration path, is able to identify the biologically reasonable choices with a much higher probability. Moreover, regardless of the \texttt{KGML-xDTD} with and without demonstration paths, we can see that they both are superior to the \texttt{MultiHop}, which illustrates that the actor-critic model structure might be more effective than LSTM for this task.

To further evaluate the performance of \texttt{KGML-xDTD} model framework in identifying biologically relevant MOA paths for drug repurposing, we present two different case studies to explore the potential repurposed drugs and their potential mechanism for two rare genetic diseases: Hemophilia B and Huntington's disease. 

\subsubsection{Case 1: Hemophilia B}\label{subsubsec:hemophilia_b}

\begin{table}
	\caption{Top 10 predicted drugs/treatments for hemophilia B (note that the drugs highlighted in red color are used in the training set).}
    \footnotesize
    \centering
	\begin{tabular}{llccllcc}
	    \toprule
		\textbf{Drug/Treatment} & \textbf{Prob.} & \textbf{Publications} \\
		\midrule
        \textcolor{red}{Eptacog Alfa (rFVIIa)} & \textcolor{red}{0.833} & \textcolor{red}{\citep{Croom_Katherine_F_2018, Minno_Giovanni_Di_2015}} \\
		\textcolor{red}{Nonacog Alfa (rFIX)} & \textcolor{red}{0.803} & \textcolor{red}{\citep{Rendo_Pablo_2015}} \\
		\textcolor{red}{Viral Vector} & \textcolor{red}{0.780} & \textcolor{red}{\citep{Driessche_Thierry_2001}} \\
		Factor VIIa & 0.748 & \citep{Roberts_Harold_R_2004, Croom_Katherine_F_2018} \\
		Recombinant FVIIa (rFVIIa) & 0.724 & \citep{Roberts_Harold_R_2004, Croom_Katherine_F_2018} \\
		Thrombin & 0.709 & \citep{Claude_Negrier_2019} \\
	  Factor IX & 0.708 & \citep{Goodeve_A_C_2015} \\
		Epicriptine & 0.702 & \\
		Hyperbaric Oxygen & 0.660 & \\
		Triamcinolone & 0.649 & \\
		\bottomrule
	\end{tabular}
\label{tab:casestudy1_res}
\end{table}

\begin{figure}[h]
\begin{center}
\includegraphics[height=6.5cm]{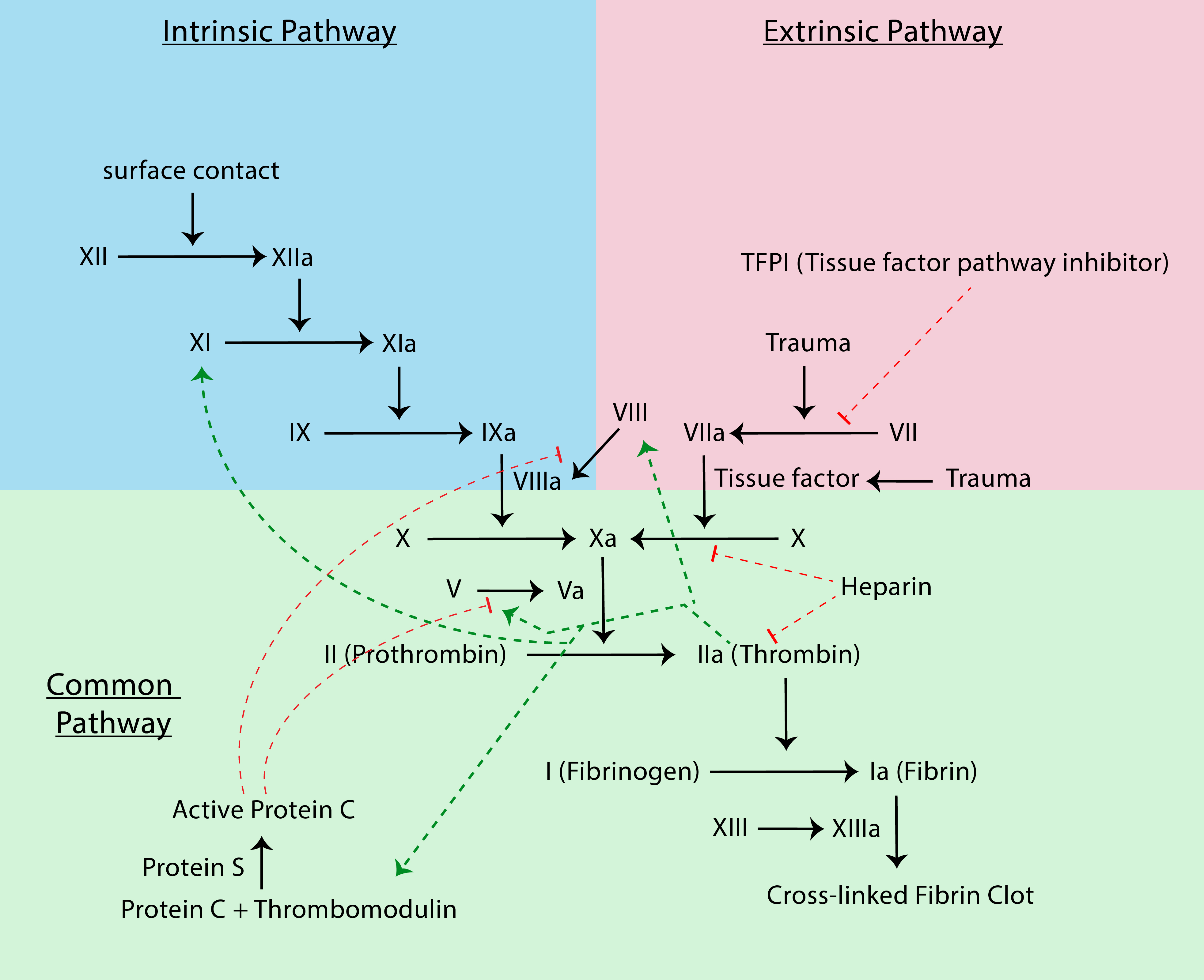}
\end{center}
\caption{Blood Coagulation Regulatory Network with arrows for molecular reactions (black), positive feedback (green), and negative feedback (red).}
\label{fig:fig4}
\end{figure}

 Hemophilia B, also known as factor IX deficiency or Christmas disease, is a rare genetic disorder that results in prolonged bleeding in patients. It is caused by mutations in the factor IX (F9) gene, which is located on the X chromosome. Table~\ref{tab:casestudy1_res} displays the top 10 drugs/treatments predicted by the \texttt{KGML-xDTD} model framework,  including both those that are used in the training set (highlighted in red) and those that are not. Besides those known drugs/treatments used in the training set, the majority of the remaining seven drugs/treatments on the list are supported by published research and have the potential to treat hemophilia B. For example, the activated human-derived coagulation factor VII (i.e., Factor VIIa) or the recombinant activated factor VII (i.e., rFVIIa) is one of the proteins that can cause blood clots as an important part of the blood coagulation regulatory network (as shown in the Figure~\ref{fig:fig4}). This protein is used as an effective inhibitor in the treatment of patients with hemophilia B \citep{Roberts_Harold_R_2004, Croom_Katherine_F_2018}. Thrombin is a key enzyme in the maintenance of normal hemostatic function. It has been reported that using thrombin as a therapeutic strategy can help prevent bleeding in patients with hemophilia \citep{Claude_Negrier_2019}. The use of recombinant factor IX therapy is a recommended treatment option for individuals with hemophilia B \citep{Goodeve_A_C_2015}. Some examples of recombinant factor IX products include BeneFIX, Rixubis, Ixinity, Alprolix Idelvion, and Rebinyn. 

\begin{figure*}[tpbh]
\begin{center}
\includegraphics[height=9cm]{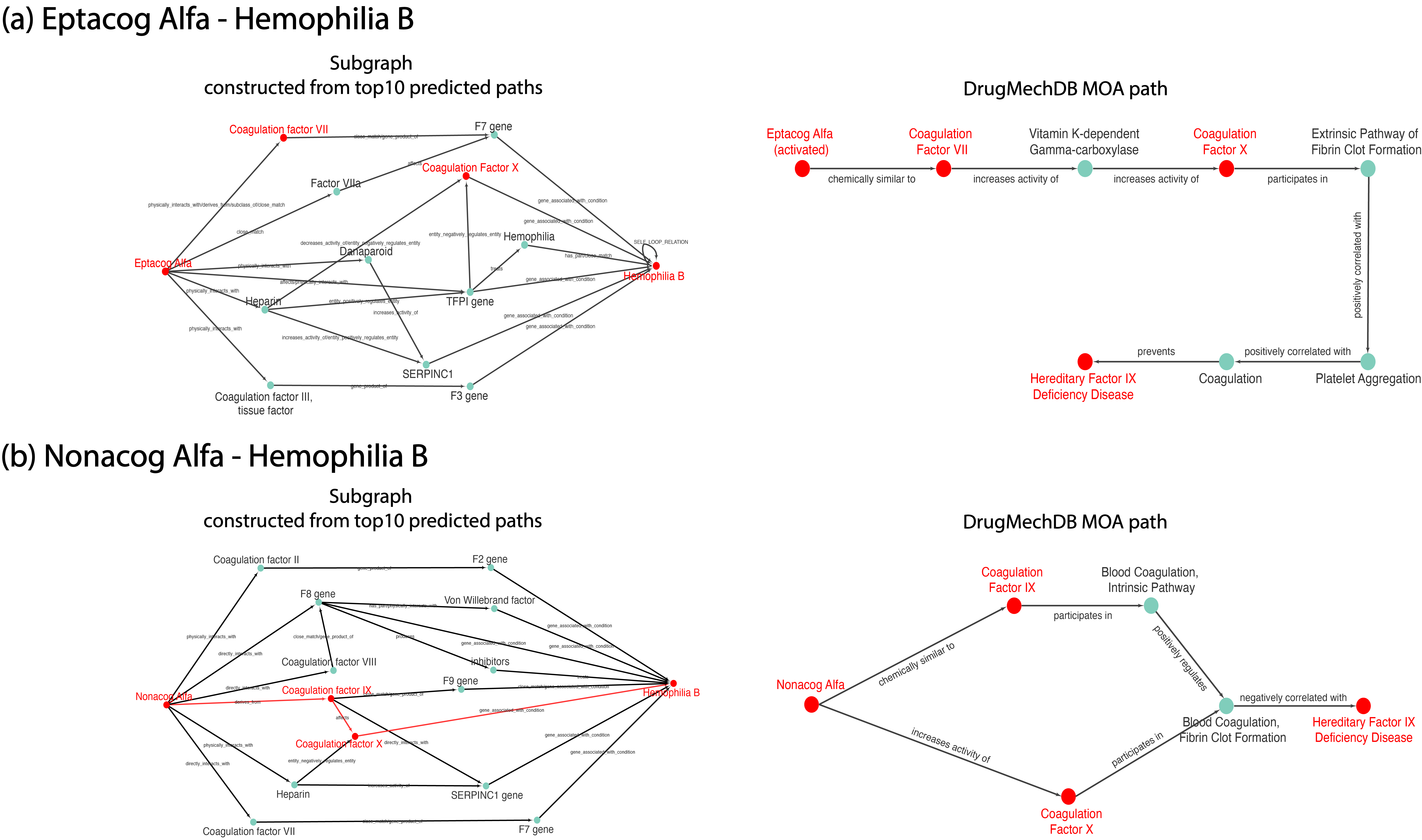}
\end{center}
\caption{Comparison between the top 10 predicted 3-hop paths (integrated into a subgraph for better visualization) and the curated DugMechDB-based MOA paths for Eptacog Alfa (a) and Nonacog Alfa (b). On the left is a network representation of the top 10 \texttt{KGML-xDTD} predicted paths. On the right are network representations of the human-curated DrugMechDB mechanism of action paths. In both, the corresponding biological entities/vertices between the top 10 predicted 3-hop paths and MOA paths are highlighted in red color. The red edges indicate a top-10-\texttt{KGML-xDTD} predicted path where all entities show up in the DugMechDB-based MOA paths.}
\label{fig:fig5}
\end{figure*}

To further assess the biological explanations of the predicted 3-hop BKG-based MOA paths for the treatment of hemophilia B, we have used the curated DrugMechDB-based MOA paths, which are not used in the model training process. DrugMechDB contains relevant MOA paths of hemophilia B treatment only for Eptacog Alfa and Nonacog Alfa. We show the comparisons between the subgraphs with the top 10 predicted 3-hop BKG-based paths and the curated DugMechDB-based MOA paths for Eptacog Alfa and Nonacog Alfa in Figure~\ref{fig:fig5}. The corresponding biological entities between the predicted paths and the curated DrugMechDB-based paths are highlighted in red color. Although the predicted paths can't exactly match the DrugMechDB-based MOA paths due to the limited path length and some missing semantic relationships in the customized biomedical knowledge graph, key biological entities (such as Coagulation Factor VII, Coagulation Factor X, and Coagulation Factor IX) that are important for the treatment of hemophilia B are present in the subgraphs of the top 10 predicted paths. As shown in Figure~\ref{fig:fig4}, the treatment of hemophilia B involves a complex molecular network of blood coagulation, and many of the coagulation factors (such as factor VII, factor VIIa, factor III, factor II, factor VIII, factor IX, and factor X) present in the subgraphs of the top 10 predicted paths are also part of this molecular network. In Appx.~\ref{sec:subgraph_moa_path_factor_viia}, we also utilize the \texttt{KGML-xDTD} model framework to predict the top 10 3-hop BKG-based paths, which can serve as biological explanations of the predicted "treats" relationship between Factor VIIa and Hemophilia B (shown in Table~\ref{tab:casestudy1_res}). This particular drug/treatment - disease pair is not included in the training set and thus can be used to indicate how \texttt{KGML-xDTD}'s MOA path predictions can contribute to the explanation of the predicted drug repurposing results. The predicted paths show molecular details akin to those in Figure~\ref{fig:fig4} for treating hemophilia B. As a result, the predicted paths by \texttt{KGML-xDTD} model framework can help identify key molecules in the real drug action regulatory network, thereby aiding in explaining drug repurposing to some extent.

\subsubsection{Case 2: Huntington's disease}\label{subsubsec:huntington}
\begin{figure*}[bh]
\begin{center}
\includegraphics[height=12cm]{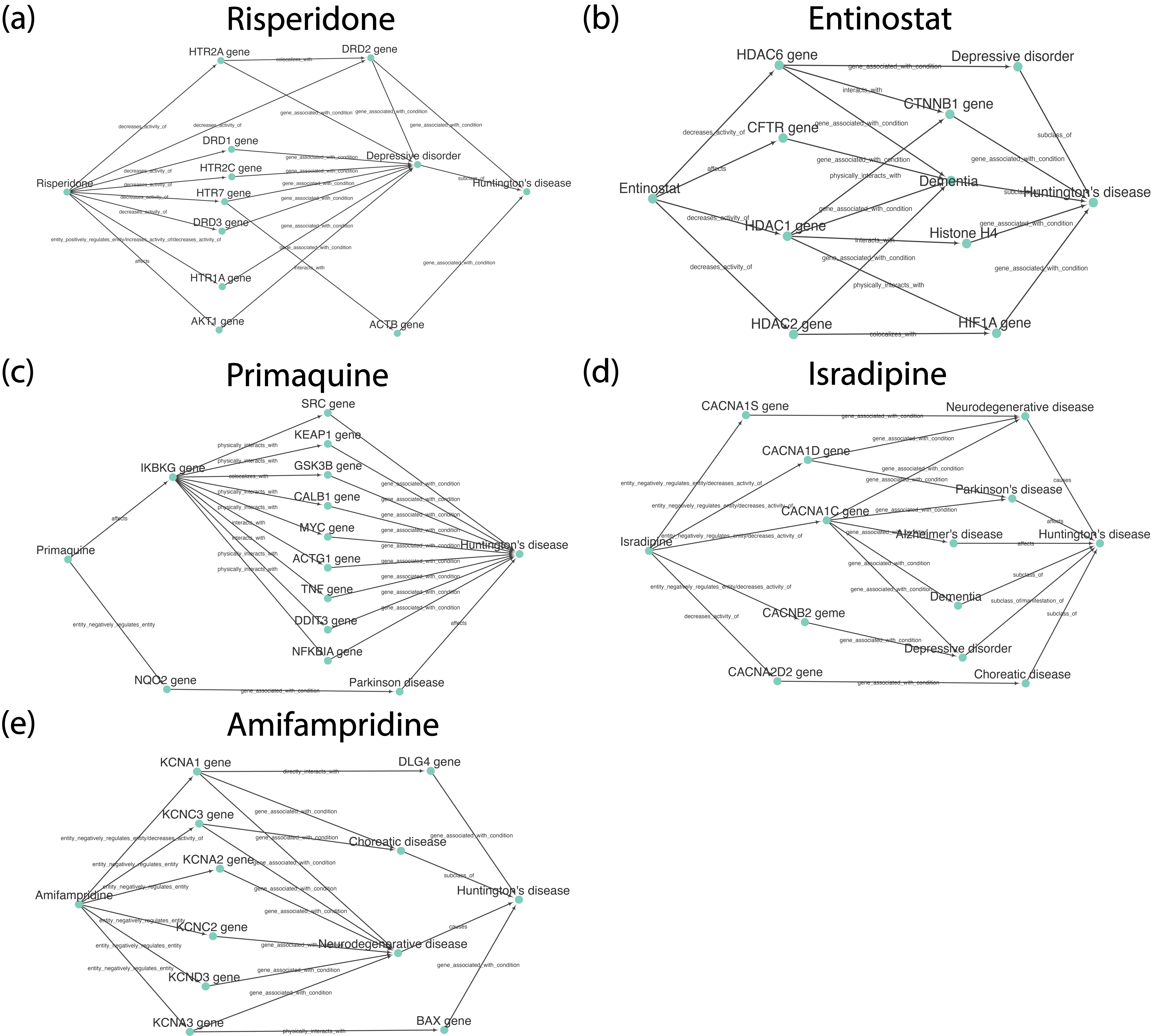}
\end{center}
\caption{Top 10 predicted 3-hop BKG-based MOA paths (integrated into subgraphs) for top 5 non-chemotherapeutic predicted drugs/treatments that are not included in the training set for Huntington's disease.}
\label{fig:fig6}
\end{figure*}

Huntington's disease (HD) is a rare neurogenetic disorder that typically occurs in midlife with symptoms of depression, uncontrolled movements, and cognitive decline. While there is currently no drug/treatment that can alter the course of HD, some drugs/treatments can be useful for the treatment of its symptoms in abnormal movements (e.g., chorea) and psychiatric phenotypes. We show ten drugs/treatments with the highest predicted probability by the \texttt{KGML-xDTD} model framework after manual processing in Table~\ref{tab:casestudy2_res}. This processing involves excluding the chemotherapeutic drugs from the predicted drug candidate list due to their potential risk of cytotoxicity to normal cells (which could lead to false positives for drug repurposing of non-cancer diseases \citep{Julien_Sourimant_2021,Gysi_Deisy_Morselli_2021}), and only presenting the top 5 results in the training set, and top 5 from the test or validation set. From this table, it can be observed that many of the top-ranked predicted drugs have been supported by publications as potential treatments for the symptoms of HD. Since there is currently no effective treatment for HD, DrugMechDB does not have a corresponding MOA path for comparison. To analyze the predicted paths by the \texttt{KGML-xDTD} model framework for the predicted non-chemotherapeutic drugs/treatments that are not included in the training set (shown in black in Table~\ref{tab:casestudy2_res}), we present their top 10 predicted paths (integrated into different subgraphs) in Figure~\ref{fig:fig6}. From these predicted paths, we can see that most of them are biologically relevant. For example, the subfigure (a) of Figure~\ref{fig:fig6} shows that Risperidone is predicted to be useful for the treatment of HD by decreasing the activity of the genes associated with the 5-Hydroxytryptamine receptor (e.g., HTR1A, HTR2A, HTR2C, HTR7) and dopamine receptor (e.g., DRD1, DRD2, DRD3) which have been proven to be involved in the pathogenesis of depressive disorders \citep{Yohn_Christine_N_2017,Delva_Nella_C_2021}. The presence of depressive symptoms is a significant characteristic of HD \citep{Coppen_Emma_M_2017}. Entinostat is predicted to have the potential to alleviate the symptoms of HD by inhibiting the functions of histone deacetylase genes such as HDAC1, HDAC2, HDAC6 (see subfigure (b) of Figure~\ref{fig:fig6}), and one of the predicted 3-hop MOA paths ("Entinostat" $\rightarrow$ "decreases activity of" $\rightarrow$ "HDAC1 gene" $\rightarrow$ "interacts with" $\rightarrow$ "Histone H4" $\rightarrow$ "gene associated with condition" $\rightarrow$ "Huntington's disease") is supported by the previous research \citep{Shukla_Surabhi_2020, Yu_In_Tag_2009}. Primaquine is predicted to act on the NQ02 gene and the IKBKG gene to potentially play a therapeutic role in neurodegenerative disease, reported in \citep{Voronin_Mikhail_V_2021, Singh_Shareen_2020}. According to the predicted MOA paths, Isradipine may have a potential therapeutic effect for HD by mainly regulating the genes of the Calcium Voltage-Gated Channel, including CACNA1S, CACNA1D, CACNA1C, CACNB2, CACNA2D2. These genes may be associated with the symptoms of HD, such as chorea, depression, and dementia \citep{Yagami_T_2012}. Lastly, Amifampridine is predicted to regulate the genes of the Potassium Voltage-Gated Channel which are potentially associated with HD \citep{Noh_Wonjun_2019}. All these examples indicate that the predicted BKG-based MOA paths can explain the mechanism of repurposed drugs to some extent.

\begin{table}
	\caption{Top 5 predicted drugs/treatments used in the training set (highlighted in red color) and the top 5 non-chemotherapeutic predicted drugs/treatments that are not in the training set for Huntington's Disease.}
    \footnotesize
    \centering
	\begin{tabular}{llccllcc}
	    \toprule
		\textbf{Drug/Treatment} & \textbf{Prob.} & \textbf{Publications} \\
		\midrule
		\textcolor{red}{Pimozide} & \textcolor{red}{0.939} & \textcolor{red}{\citep{Arena_R_1980, Videnovic_Aleksandar_2013}} \\
        \textcolor{red}{Therapeutic Agent} & \textcolor{red}{0.939} &  \\
		\textcolor{red}{Olanzapine} & \textcolor{red}{0.938} & \textcolor{red}{\citep{Paleacu_D_2002,Squitieri_F_2001}} \\
		\textcolor{red}{Riluzole} & \textcolor{red}{0.935} & \textcolor{red}{\citep{Group_Huntington_Study_2003}} \\
		\textcolor{red}{Antipsychotic Agent} & \textcolor{red}{0.932} & \textcolor{red}{\citep{Unti_E_2017}} \\
		Risperidone & 0.893 & \citep{Coppen_Emma_M_2017,Duff_Kevin_2008} \\
		Entinostat & 0.888 & \citep{Shukla_Surabhi_2020} \\
		Primaquine & 0.887 & \\
		Isradipine & 0.884 & \citep{Miranda_Artur_S_2019} \\
		Amifampridine & 0.882 & \\
		\bottomrule
	\end{tabular}
\label{tab:casestudy2_res}
\end{table}

\section{Discussion}
In this work, we propose \texttt{KGML-xDTD}, a two-module, knowledge graph-based machine learning framework that not only predicts the treatment probabilities between drugs/compounds and diseases but also provides biological explanations for these predictions through the predicted paths in a massive biomedical knowledge graph with comprehensive biomedical data sources as potential mechanisms of action. This framework can assist medical researchers in quickly identifying the potential drug/compound-disease pairs that might have a treatment relationship, which can accelerate the process of drug discovery for emerging diseases. Additionally, by leveraging the KG-based paths predicted by the framework, medical professionals (e.g., doctors and licensed medical practitioners) can straightforwardly assess the accuracy of the predictions via the predicted MOAs, which can help to reduce false positives that may be produced by the "black-box" operation of traditional machine learning models.

Although previous research \citep{Himmelstein_Daniel_Scott_2017, Zhang_Rui_2021, Vassilis_N_Ioannidis_2020} has applied a variety of models to the task of drug repurposing using biomedical knowledge graphs (BKGs), these approaches are implemented in the small-scale BKGs and many do not scale to larger graphs. As biotechnology advances and the volume of data in biomedical databases increases, BKGs are becoming larger and more complex. In our comparison with state-of-the-art KG-based models for drug repurposing, we find that the \texttt{KGML-xDTD} model had higher accuracy with lower false positives when applied to a massive and complex biomedical knowledge graph \emph{RTX-KG2c}.
By evaluating the predicted paths with DrugMechDB and two case studies, we show that the model can capture some key biological entities involved in real drug action regulatory networks.

It is widely acknowledged that drug repurposing is one of the most challenging problems in biomedicine, and current AI techniques are still in the early stages of addressing it. Many other AI models, such as those based on chemical structure, drug-target interactions, and drug perturbations of gene expression, are developed for solving this goal. They may offer more accurate predictions but also have limitations in terms of cost and the availability of samples for specific diseases. Biomedical knowledge graph (BKG)-based machine learning models, such as the \texttt{KGML-xDTD} model, offer a cost- and time-efficient alternative due to the large volume of biomedical knowledge stored in public databases and publications. The \texttt{KGML-xDTD} model framework is not intended to replace or beat these models, but rather provides a complementary approach that leverages emerging knowledge graphs for drug repurposing.

Future work to further enhance the \texttt{KGML-xDTD} model framework might include extending the predicted paths for more specific explanations, and considering the negative drug-disease pairs so that the model can explain why certain drugs are harmful to diseases. 

\clearpage
\bibliographystyle{unsrtnat}
\bibliography{main}

\clearpage
\appendix

\section{Section S1. Biomedical Knowledge Graph RTX-KG2c Pre-processing}\label{sec:bkg_preprocessing}

The dataset of \emph{RTX-KG2c} (v2.7.3) \citep{Wood_E_C_2022} is accessed via \href{https://github.com/RTXteam/RTX-KG2}{https://github.com/RTXteam/RTX-KG2}. We pre-process the raw data of \emph{RTX-KG2c} by the following four principles:
\begin{enumerate}
  \item Since we are mainly interested in the categories relevant to drug mechanisms of action (MOAs), we exclude the nodes with categories that are not expected to be useful for drug repurposing explanation (e.g., ``GeographicaLocation", ``Device", ``InformationResource").
  \item One of data sources used in the \emph{RTX-KG2c} is the Semantic MEDLINE Database (SemMedDB) \citep{Kilicoglu_Halil_2012}, one of the most widely used NLP-derived biomedical knowledge sources, that has been found recently \citep{Cong_Qing_2018} to contain some inconsistent relations due to the immature NLP techniques even though it contains many latest-found relations (e.g., the relations with Covid-19). To improve the quality of SemMedDB-based edges, we filter out some edges based on the following criteria:
   \begin{itemize}
     \item Each remaining SemMedDB-based edge must be supported by at least 10 publications.
     \item The PubMed-publication-based NGD score \citep{Cilibrasi_R_L_2007} (defined in Equation 1 in the main text) of the two end nodes should be higher than 0.6.
   \end{itemize}
  \item The \emph{RTX-KG2c} is a multigraph that allows multiple edges to connect between two nodes. These edges present different relations (a.k.a. predicates) between the nodes, and follow the predicate hierarchy \footnote{Visualization of hierarchical predicates in the Biolink Model v2.1.0 \url{http://tree-viz-biolink.herokuapp.com/predicates/2.1.0}} used in the Biolink model \citep{unni2022biolink}. As a result, the \emph{RTX-KG2c} contains some hierarchically redundant edges between two nodes. To simplify the process of path finding for the downstream MOA predictions, we only reserve the ``leaf" predicates (i.e., the most specific predicates) in the Biolink semantic relation hierarchy if there are hierarchically associated edges between two nodes. For example, if there are edges between two nodes representing the predicates ``affected by", ``entity regulated by entity" and ``entity positively regulated by entity", we would remove the ``affected by" and ``entity regulated by entity" edges because they are more general (i.e., ``ancestor") predicates in the hierarchy compared to ``entity positively regulated by entity". Removing those more general predicates does not affect the interpretability of the paths because the “leaf” predicates contain more precise semantic information.
  \item To avoid the training information leakage, we exclude all existing edges that directly connect the potential drug nodes (nodes with categories “Drug” or “SmallMolecule”) to potential disease nodes (nodes with categories ``Disease", ``PhenotypicFeature", ``BehavioralFeature" or “DiseaseOrPhenotypicFeature”) in the \emph{RTX-KG2c}.
\end{enumerate}
After applying the pre-processing steps to \emph{RTX-KG2c} (v2.7.3) described above, the post-processed/customized biomedical knowledge graph (BKG) consists of 3,659,165 nodes with 33 distinct categories, and 18,291,237 edges with 74 distinct types.

\section{Section S2. Implementation Details of \texttt{KGML-xDTD} Model Framework}\label{sec:implemention_KGML_xDTD}
\subsection{Drug Repurposing Prediction Module}

In the drug repurposing prediction module, we utilize the source code \footnote{\url{https://github.com/williamleif/GraphSAGE}} provided by \citet{Hamilton_William_L_2017} to train unsupervised GraphSAGE embeddings with its “big” mean-based aggregator and two hidden layers of dimensions [256, 256]. For the random walk setting, we performed 10 walks each with length of 100. As for other parameters, the number of epochs is set to 10, the neighbor sampling size of each layer is 96, the learning rate is 0.001, the batch size is 256, and the maximum number of iterations per epoch is 10,000. Instead of using the default identity embeddings as the initial features, we use node attribute embeddings generated by the pre-trained PubMedBert model \citep{Gu_Yu_2022} with concatenation of the node's name and category, and then reduce their dimensions to 100 using Principal Component Analysis (PCA). The final output output embedding vector for each node has a dimension of 512. With these GraphSAGE embedding vectors, we concatenate the embedding vectors of each drug-disease pair in the training set as input features and use the \emph{RandomForestClassifier} function of scikit-learn (v1.0) python package to train a Random Forest model. We run a grid search using the \emph{GridSearchCV} function to determine the optimal parameter set for the Random Forest model from a range of depths \{5, 10, 15, 20, 25, 30, 35\} and number of trees \{500, 1000, 1500, 2000\}. The best parameter set for the Random Forest model uses the maximum depth $max\_depth=35$ and the number of trees $n\_estimators=2000$. 

\subsection{Mechanism of Action (MOA) Prediction Module}
In the mechanism of action prediction module, we design a reinforcement learning model following the model framework of \citet{Zhao_Kangzhi_2020} for drug repurposing purpose (see Sec. ``Mechanism of Action (MOA) Prediction" in the main text). We set the state history length $K=2$ and the maximum length of path $T=3$. In order to make the customized biomedical knowledge graph (described in Sec.~\ref{sec:bkg_preprocessing} above) enable training on a 48GB Quadro RTX 8000 GPU, we prune the action space of each node to a maximum size of 3,000 based on the PageRank score \citep{Lawrence_Page_1999} (calculated by the \emph{pagerank} function of NetworkX (v2.7.1) python package). We set the dimensions of all lookup matrices used within actor, critic, meta-path discriminator, and path discriminator networks to 100. The dimensions of hidden layers of the actor network and critic network are both set to 512. We set the dimensions of hidden layers of the path discriminator with [512, 512] and used the dimension set of [512, 256] for the hidden layers of the meta-path discriminator. We use Xavier initialization \citep{Glorot_Xavier_2010} for the embeddings of all lookup matrices and the network layers. The weight of the path discriminator reward ${\alpha}_{p}$ is set to $0.006$ while the meta-factor of the path discriminator reward $a_{m}=0.012$. We respectively assign 0.99 and 0.005 to the decaying coefficient $\gamma$ of $R_{e,T}$ and the weight $\alpha$ of entropy term. We optimize all networks using the Adam optimization algorithm \citep{Diederik_P_Kingma_2015} with a learning rate of 0.0005. The mini-batch size is set to 32 with a path rollout of 35. The dropout rates of all subnetworks are set to 0.3 and the action dropout rate is set to 0.5.

\section{Section S3. Implementation Details of Demonstration Path Extraction}\label{sec:implemention_dpe}

The demonstration paths are a set of multi-hop BKG-based paths that can be used to guide the agent in the reinforcement learning model of \texttt{KGML-xDTD} model framework to find biologically reasonable BKG-based MOA paths. It can be formulated as $P^{k}=\{p_{s,t}^{k} | v_{s} \in \mathcal{V}^{\mathrm{drug}}; v_{t} \in \mathcal{V}^{\mathrm{disease}}\}$ where $p_{s,t}^{k}$ is a multi-hop demonstration path with maximum path length $k$ starting from a drug node $v_{s}$ and ending at a disease node $v_{t}$. Given a potential drug node and a potential disease node (defined in Sec.~\ref{sec:bkg_preprocessing} above), the number of paths in the customized biomedical knowledge graph between them grows exponentially as the value of $k$ increases. Therefore, we set $k=3$ to guarantee that the agent can find biologically meaningful predicted MOA paths in a reasonable amount of time. To extract reasonable demonstration paths from the customized biomedical knowledge graph, we use the known drug-target interactions collected from two curated biomedical data sources (e.g., DrugBank (v5.1) and Molecular Data Provider (v1.2)\footnote{\url{https://github.com/NCATSTranslator/Translator-All/wiki/Molecular-Data-Provider}}), as well as the PubMed-publication-based Normalized Google Distance (NGD) (see Equation 1 in the main text). A demonstration path extracted from the customized biomedical knowledge graph must satisfy the following two requirements: 
\begin{enumerate}
  \item The edge connecting the drug node to the first intermediate node in a demonstration path must be supported by DrugBank or Molecular Data Provider as a known drug-target interaction, and must also have a Normalized Google Distance (NGD) score of 0.6 or lower. 
  \item The edge connecting the second intermediate node to the disease node in a demonstration path should have a Normalized Google Distance (NGD) score of 0.6 or lower.
\end{enumerate}
Only the true positive pairs (see Table 1 in the main text) collected from four human-curated and NLP-derived training datasets (described in Sec. ``Data Sources for Model Training" in the main text) are used to extract demonstration paths. We also filter out any true positive pairs that are not reachable from a drug node to its corresponding disease node within maximum of 3 hops in the customized biomedical knowledge graph. Out of these 21,437 true positive pairs, 8,495 are able to find at least one demonstration path that meet the requirement. We finally find 396,705 demonstration paths for 8,495 true positive drug-disease pairs which are used for reinforcement learning training. 

\section{Section S4. Implementation Details of Baseline Models}\label{sec:implemention_baselines}

We use the OpenKE library \footnote{\url{https://github.com/thunlp/OpenKE}} to implement the models \texttt{TransE}, \texttt{TransR}, \texttt{RotatE}, \texttt{DistMult}, \texttt{ComplEx}, \texttt{ANALOGY}, and \texttt{SimpLE} with their default parameter settings (we adjusted the hyperparameters for some models to ensure we can finish the training on GPUs for a reasonable time). Table \ref{tab:openke} shows the detailed hyperparameter setting of these baseline model. For other baseline models, we use the PyTorch Geometric \footnote{\url{https://github.com/pyg-team/pytorch_geometric}} framework to implement the \texttt{GAT} and \texttt{GraphSAGE-link} model and use the same GraphSAGE embeddings (mentioned above) with scikit-learn (v1.0) python package to implement the \texttt{GraphSAGE+logistic}, \texttt{GraphSAGE+SVM} and \texttt{2-class GraphSAGE+RF} models with the grid-search-based optimal parameter settings.

\begin{table}[h]
\centering
\caption{Hyperparameters used for Baseline Models.}\label{tab:openke}
{
	\begin{tabular}{lccccc}
	    \toprule
		\textbf{Model} & \textbf{Hidden Dim.} & \textbf{Num. Epochs} & \textbf{Batch Size} & \textbf{Learning Rate} & \textbf{Optimizer} \\
		\midrule
		TransE & 100 & 10000 &1000 &1 & SGD \\
		TransR & 50 & 2000 & 1000 &1 &SGD \\
		RotatE & 30 & 2000 & 1000 &2e-5 &Adam \\
		DistMult & 100 &10000 &1000 &0.5& Adagrad \\
		ComplEx & 50 & 2000&500 & 0.5&Adagrad\\
		ANALOGY & 20 &2000 &500& 0.5&Adagrad \\
		SimpLE & 100 &2000 &500& 0.5&Adagrad \\
		\bottomrule
	\end{tabular}
}
\end{table}

For the implementation of the \texttt{MultiHop} model, we use the source code\footnote{\url{https://github.com/salesforce/MultiHopKG}} provided by \citet{Lin_Xi_Victoria_2018} and modify its reward function by using our defined reward shaping strategy (described in Sec. "Adversarial Actor-critic Reinforcement Learning" in the main text). We set all its parameters the same as the reinforcement learning model in the \texttt{KGML-xDTD} model framework if they are available otherwise we use the default parameters.

\section*{Section S5. A subgraph with the top 10 predicted 3-hop MOA paths for Factor VIIa – Hemophilia B pair}\label{sec:subgraph_moa_path_factor_viia}

We utilize the \texttt{KGML-xDTD} model framework to predict the top 10 3-hop BKG-based paths, serving as biological explanations of the predicted "treats" relationship between Factor VIIa and Hemophilia B (shown in Table 4 for case study 1 in the main text). This particular drug/treatment - disease pair is not used in the training set and thus can be used to indicate how KGML-xDTD’s MOA path predictions can contribute to the explanation of the predicted drug repurposing results. The predicted paths show similar molecular details as those in Figure~\ref{fig:fig4} in the main text for treating hemophilia B. The path that we highlighted in red is the one in which all nodes can align with the key molecules in the real drug action regulatory network shown in Figure~\ref{fig:fig4}.

\begin{figure}[tpbh]
\begin{center}
\includegraphics[height=8cm]{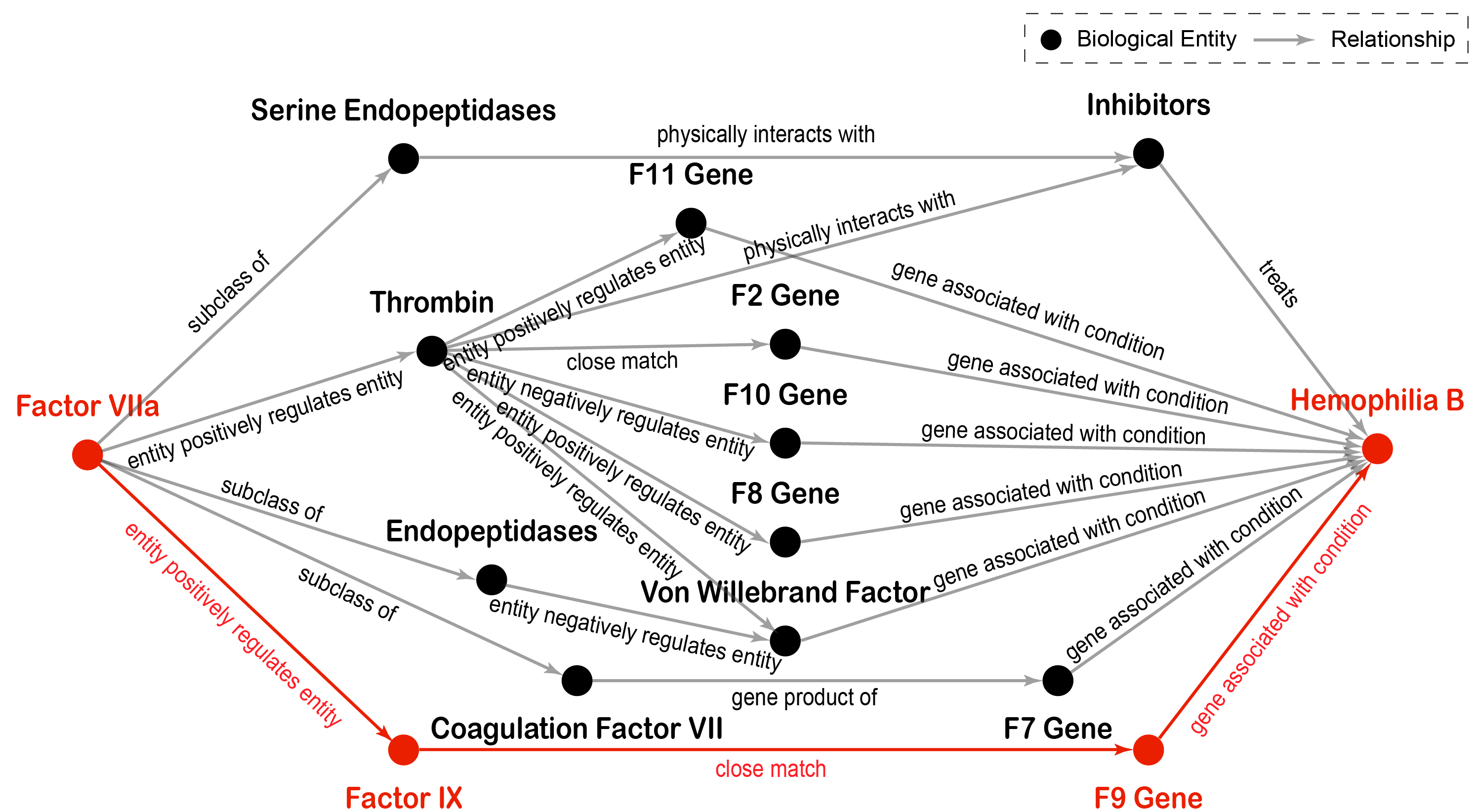}
\end{center}
\caption{Top 10 predicted 3-hop paths (integrated into a graph for better visualization) generated by the \texttt{KGML-xDTD} model framework serve as biological explanations of the predicted "treats" relationship between Factor VIIa and Hemophilia B. The path highlighted with red is the one in
which all nodes can align with the key molecules in the real drug action regulatory network shown in Figure~\ref{fig:fig4} in the main text.}
\label{suppl_fig:fig1}
\end{figure}

\end{document}